\documentclass[]{spie}  

\title{The Effects of Binary Reference Stars on JWST NIRCam Coronagraphy}

\author[a]{Klaus Subbotina Stephenson}
\author[b]{Aarynn L. Carter}
\author[a]{Andrew Skemer}
\affil[a]{Department of Astronomy \& Astrophysics, University of California, Santa Cruz, 1156 High St, Santa Cruz, CA 95064, USA}
\affil[b]{Space Telescope Science Institute, 3700 San Martin Drive, Baltimore, MD 21218, USA}

\usepackage{amsmath,amsfonts,amssymb}
\usepackage{graphicx}

\usepackage{float}
\usepackage{placeins}
\usepackage{tabularx}
\usepackage{xcolor}
\usepackage{listings}
\usepackage{xparse}
\NewDocumentCommand{\codeword}{v}{%
\texttt{\textcolor{blue}{#1}}%
}

\usepackage[colorlinks=true, allcolors=blue]{hyperref}

\begin{document} 
\maketitle

\begin{abstract}
Direct imaging campaigns executed with James Webb Space Telescope (JWST) will enable the study of the faintest observable exoplanets yet. To assist observers in the JWST proposal process, we present an in-depth exploration of the effects of moderate contrast binary (visual and physical) systems on JWST NIRCam coronagraphic Reference Differential Imaging (RDI) methods. All work in this paper is based on simulation data generated using the python package \codeword{PanCAKE}, in addition to a variety of custom scripts which we have made publicly available on GitHub. Presenting both contrast curves and more involved `heatmaps' of sensitivity loss, we present quantifiable measurements for how a binary companion will impact contrast both totally and locally, as a function of magnitude, separation, and position angle. Observers can use results in this work to estimate the impact of a known binary, and in some cases will find that PSF subtraction can still be reliably performed. We have found several scenarios where JWST NIRCam coronagraphic RDI PSF subtraction can be viably performed using a binary reference, and make several suggestions. The brightest binary companions analyzed, with a relative brightness of $10^{-3}$, resulted in the worst local sensitivity loss of $3.02$ units of magnitude. The faintest binary companions looked at, $10^{-6}$ relative brightness, have almost no effect on local sensitivity. Changing position angle impacts sensitivity loss by $0.5-0.3$ depending on companion flux. Binary companion separation considerations should be on a case-by-case science goal basis. This work also discusses the trade space for these suggestions in detail.
\end{abstract}


\section{INTRODUCTION}
\label{sec:intro}  
Direct imaging methods in exoplanetary science have gained massive momentum within the community since their development in the early 2000s (\cite{2023ASPC..534..799C, 2016PASP..128j2001B}). Several present observatories and instruments, including the Gemini Planet Imager, ESO's VLT/SPHERE instrument, and now James Webb Space Telescope (JWST) support direct imaging methods, supplying a host of opportunities for high contrast exoplanetary science (e.g., \cite{2014PNAS..11112661M, 2017A&A...605L...9C}). The direct imaging of exoplanetary systems can yield information on bulk planetary properties, disk structure, and potentially evolutionary and formation histories (\cite{2008Sci...322.1348M, 2024AJ....167..182C, 2023NatAs...7.1208W}). Additionally, direct imaging has provided a bulk of current characterizations and constraints on the properties of exoplanetary atmospheres (e.g., \cite{2013Sci...339.1398K}, \cite{2023ASPC..534..799C}, \cite{2017AJ....154..218N}).

JWST marks the start of a new revolutionary phase of exoplanet science. JWST hosts the largest primary mirror of any space telescope to date, permitting sensitivity to fainter objects than those afforded by the Hubble and Spitzer space telescopes. JWST's broad infrared wavelength coverage and excellent pointing stability enables unprecedented contrast sensitivity at wavelengths beyond 2 $\mu$m (\cite{Beichman_2010}), and JWST Near Infrared Camera's (NIRCam) sensitivity at 3-5 $\mu$m permits the study of low mass planets not detectable from ground based observatories (\cite{2018arXiv180303730B,2005SPIE.5905..185G}). The ability to directly image down to to the lowest achievable exoplanet masses of 0.1 $M_{\mathrm{Jup}}$ at large separations beyond 10 au is unique to JWST, setting it apart from current observatories (\cite{2021MNRAS.501.1999C}).

Nevertheless, direct imaging remains challenging due to the difference in magnitude between the emitted stellar and planetary flux. The overwhelming luminosity of a host star effectively drowns out any emitted planet light and must be carefully handled to improve the likelihood of identifying planets. To do this, one can use a coronagraph, consisting of an occulting mask and Lyot stop. Better described as an opaque focal plane mask, occulting masks are placed optically conjugated to the telescope pupil. This mask will block or diffract on-axis starlight away from the pupil plane, leaving off-axis planet light to pass through the mask undisturbed (\cite{2024CRPhy..24S.133G}). JWST offers two shapes of occulting masks, two tapered bar variations and three round variations. JWST implements occulting masks with Lyot stops, which are separate from occulting masks. Lyot stops, also utilized to suppress refracted light, are located at the focus of the entrance pupil over top each of JWST's primary mirror segments, excluding segments underneath spider arms (\cite{2010SPIE.7731E..3JK}). 

A coronagraph centered over a target star will suppress the majority of on-axis starlight, however, some remnant scattered and diffracted light from this occulted star remains (\cite{2012AAS...21915502T}). Planetary flux must be further revealed through the use of post processing techniques that rid the science images of the residual starlight leftover from coronagraphic observation. The subtraction of this residual Point-Spread Function (PSF) is necessary for reaching the highest achievable Signal-to-Noise Ratio (SNR) for coronagraphic observations, and attaining sensitivity to binary companions, such as exoplanets, at short angular separations (\cite{2019AJ....158...36G}). Both JWST's Mid-Infrared Instrument (MIRI) and NIRCam instruments have coronagraphic capabilities, but in this paper we focus on NIRCam coronagraphy.

\subsection{Angular Differential Imaging}
Angular Differential Imaging (ADI) is one of the PSF subtraction methods supported by JWST, and requires the sequential observation of the target star at different observatory roll angles. This process is different for ground- and space-based observatories. Ground-based observatories achieve roll angle images by turning the imaging instrument's field derotator off, keeping optics aligned while permitting the instrument's Field Of View (FOV) to rotate with respect to the telescope. For space observatories such as JWST, the observatory and all of its instruments physically roll on its present axis. An appropriately selected image from the observation sequence is selected for use as a subtraction to other images from the same sequence. This subtraction of two images from the same observational sequence is called `roll deconvolution' (\cite{1985SPIE..556..270M}). After a reference PSF has been subtracted from all images for a sequence, frames are rotated to align to a single FOV. A median combination of all the frames are combined for a final ADI subtracted image. 

ADI imaging has the advantage of minimal wavefront drifts between PSFs due to the reference PSF being taken from target observations. Furthermore, ADI imaging is an efficient choice for proposals with limited time, as there is no need to dedicate time for imaging a separate reference PSF source. However, there are several known features that can make ADI undesirable for JWST exoplanetary imaging observations. Small roll angles increase the ADI self-subtraction factor, a phenomenon dependent on instrumentation FOV position angle. For a given roll angle, companions at small separations make little movement on a detector between rolls. Larger separation companions move further distances on the same detector in the same time frame as more closely separated objects (\cite{2016SPIE.9909E..7QC}). Therefore, objects at shorter separations suffer from more self-subtraction in comparison to objects at larger separations under the same roll angle. JWST has a small potential roll angle of $\sim$10 degrees, limiting performance at small angular separations where faint companions are more likely to exist (\cite{2022SPIE12180E..0SH}). JWST's small roll angle has also been described negatively in several JWST imaging campaigns, such as distorted disk appearance in \cite{2024AJ....167..181W}, and severe self-subtraction of HL Tau Disk features in \cite{2024AJ....167..183M}

\subsection{Reference Differential Imaging}
Reference Differential Imaging (RDI) serves as a powerful alternative to ADI methods. Rather than taking a selection of observations at different roll angles, a telescope will slew between a science target and a reference target, containing an on-axis source of similar magnitude and spectral type to the science target on-axis host star. These are separate observations from the science observations and can be conducted in between science frames, or after all science images have been taken. RDI PSF subtractions are then performed with the reference PSF, rather than through self-subtraction. RDI imaging provides superior sensitivities at small angular separations, where ADI imaging suffers from a self-subtraction factor, and in some cases improves detection sensitivity by up to a factor of 5 with respect to ADI imaging (\cite{2019AJ....157..118R, 2024AJ....167..183M}). This is especially useful for observers wanting to image companions at the closest angular separations (\cite{2022SPIE12180E..3QG}). Additionally, the timescale for telescope slewing between science and reference targets is insignificant when considering wavefront stability. During JWST Cycle 1, wavefront variations between observations were on the order 10nm rms, compared to a measurement uncertainty of 7nm rms (\cite{2023PASP..135d8001R}). In relation to campaigns concerned with disk imaging, RDI methods can reveal more features and provide a more robust recovery disk morphology (\cite{2022A&A...666A..32X}). It is for these reasons listed that multiple exoplanet programs from JWST Cycle 3 will be utilizing RDI imaging to carry out direct imaging campaigns, such as \cite{2024jwst.prop.6122B} and \cite{2024jwst.prop.5835C}.

There are several reference source requirements that should be met for a successful JWST RDI imaging proposal. These requirements leave little flexibility for reference candidates, and can be particularly hindering to a proposal if a preferred reference source (or shortlist of references sources) does not meet JWST RDI standards. The major constraints, in no particular order, are as follows:
\begin{enumerate}
  \item Spectral Type -- The spectral type of the reference PSF must match the on-axis science target as closely as possible. Otherwise, a spectral mismatch degrades PSF subtraction due to differences in the reference and science target spectral energy distribution (\cite{2000Msngr..99...31M}). \footnote{The various effects of spectral mismatches in relation to sensitivity loss has been well studied and can be found on the official JWST User Documentation \cite{JWST2024}}
  \item Magnitude -- Brighter reference sources are more desirable, as integration times lower as a function of magnitude. Dimmer reference sources need long exposure times to achieve a similar SNR, forcing observers to spend less time on the science target and/or request more telescope time. We also note that, where possible, it may be desirable to reach a similar fraction of detector full well for a brighter reference to mitigate the PSF modifying influence of the ``brighter-fatter" effect (\cite{2024ApJ...963L...2S}). 
  \item Slew limitations -- Large slews from science to reference scenes take more time, increasing the likelihood of wavefront drift. This can also have consequences for proposals that need to consider time constraints. 
  \item Singular PSF -- Ideally, the reference PSF is isolated so that no additional off-axis contaminating sources are present when performing the PSF subtraction. This includes background objects as well as foreground objects within the reference object/system.
\end{enumerate}

The perfect RDI subtraction reference PSF is as alike to the on-axis science target as possible; extremely similar in spectral type and bright in magnitude. It can be somewhat straightforward to find a closely-separated single source star with a matching spectral type to act as a reference PSF for JWST RDI subtraction. However, additional reference source constraints on magnitude and binarity could be overly restrictive, and potentially yield no `perfect' reference for any single direct imaging proposal. In such a situation, a compromise needs to be made, such as sacrificing sensitivity by expanding potential choices to larger slews, or allowing for some level of off-axis contamination. Alternatively, one could suffer higher exposure times, and select a comparably faint reference source.

\subsection{Reference Targets with binary companions}
Presently, there is no exploration into JWST RDI subtractions performed with multi-PSF reference sources. When single source reference candidates fall short of JWST proposal requirements, or in the scenario that a single source reference candidate meets requirements but the reference is not desirable for a clean subtraction (reference is too faint, not a similar spectral type, etc.), observers may turn to multi-PSF references. When vetting observations for chosen JWST reference sources discover a binary companion, observers may even be forced to use a binary reference for their RDI subtraction. It is for these scenarios that a multi-PSF, or moderate contrast binary, may provide as a suitable alternative (or be required). These multi-PSF reference targets will be referred to as binary/ies throughout the paper, but the definition here also includes reference targets of single-PSF objects with additional an off-axis PSF present from a background source. 

Binary references can contain the same attributes as single-source references, or in many cases, even more desirable characteristics such as a closer match in spectral type or brighter magnitude. Another motivation for exploring binaries as references is the strategy of utilizing science targets as references for a campaign's other science targets. There is potential for using a known binary science target as a reference for different science targets' RDI subtraction. This would allow JWST users to eliminate time spent observing solely reference targets, and broaden the list of already approved science targets that can be used as references.

The trade off of using a binary companion comes down to the undesirable binary companion PSF. RDI subtractions performed with a binary system will be affected by this additional binary companion. RDI subtraction is traditionally conducted with an isolated, single source reference PSF, with the final output ideally featuring no other light than the object of interest. Using a binary reference breaks this tradition, with off-axis reference source light never being subtracted away; having a multi-PSF reference essentially creates an additional pollutant of unwanted flux in the final RDI subtracted image. Due to the previously lacking demand to explore this option for RDI imaging methods, binary companions have not been seriously evaluated as reference sources. This paper presents several case studies of how binary systems, and specific qualities of binary systems, affect JWST NIRCam coronagraphic RDI subtraction methods.

\vspace{3mm}

In the previous sections we have explained the setup needed to understand and motivate an exploration of binary systems as reference sources for NIRCam coronagraphy. The following section describes simulation methods, with a general overview of packages/software used in Section \ref{sec:01}. Section \ref{sec:02} discusses the use of the python package \codeword{PanCAKE}, Section \ref{sec:2.5} covers contrast curves. Following contrast curves we delve into a more involved approach for addressing the perceived effects of off-axis reference sources, in Sections \ref{sec:03}, \ref{sec:04}, and \ref{sec:4.5}. Results are presented and discussed in Section \ref{discussion}, and recommendations are summarized in Section \ref{conclusion}.

\section{JWST simulations}
\subsection{Software and Packages Overview} \label{sec:01}
The PANdeia Coronagraphy Advanced Kit for Extractions, or \codeword{PanCAKE}, is an advanced simulation tool for JWST coronagraphy (\cite{2021SPIE11823E..0HC}). \codeword{PanCAKE} is a `structural wrapper' that expands upon \codeword{Pandeia}, the official JWST exposure time calculator, to have more accurate coronagraphic simulations and friendlier user functionality. 
Mimicking input parameters required by JWST proposals, users can specify observations of `target' (science) and `reference' scenes, each consisting of a collection of specified astrophysical objects. The astrophysical scene construction requires details like spectral type, magnitude, position of an object offset from the scene's center, and location with respect to the center of the image. By specifying the instrument used, exposure filters, roll angles, grid dithers, coronagraph, and exposure parameters, \codeword{PanCAKE} can simulate observation sequences of the target and reference scenes. \codeword{PanCAKE} provides additional functionality beyond this basic description, and more information can be found in the publication \cite{2021SPIE11823E..0HC}, or the online documentation found \cite{here}. The numerous \codeword{PanCAKE} outputs offered include information such as contrast curves, PSF subtractions, and target-reference observations. \codeword{PanCAKE} outputs utilize \codeword{WebbPSF} and/or \codeword{pyKLIP} (\cite{2012SPIE.8442E..3DP}, \cite{2015ascl.soft06001W}).

For the work completed in this paper, several custom scripts, as well as specific \codeword{PanCAKE} simulations, were created to examine the effects of different moderate contrast binary characteristics on JWST NIRCam coronagraphy. Custom scripts and \codeword{PanCAKE} simulations will be described as relevant in the sections following, but can also be found on our GitHub page \cite{klaus-stephenson_2024}.

\subsection{PanCAKE Scene Setup} \label{sec:02}
\begin{figure}[ht!]
\includegraphics[width=\textwidth]{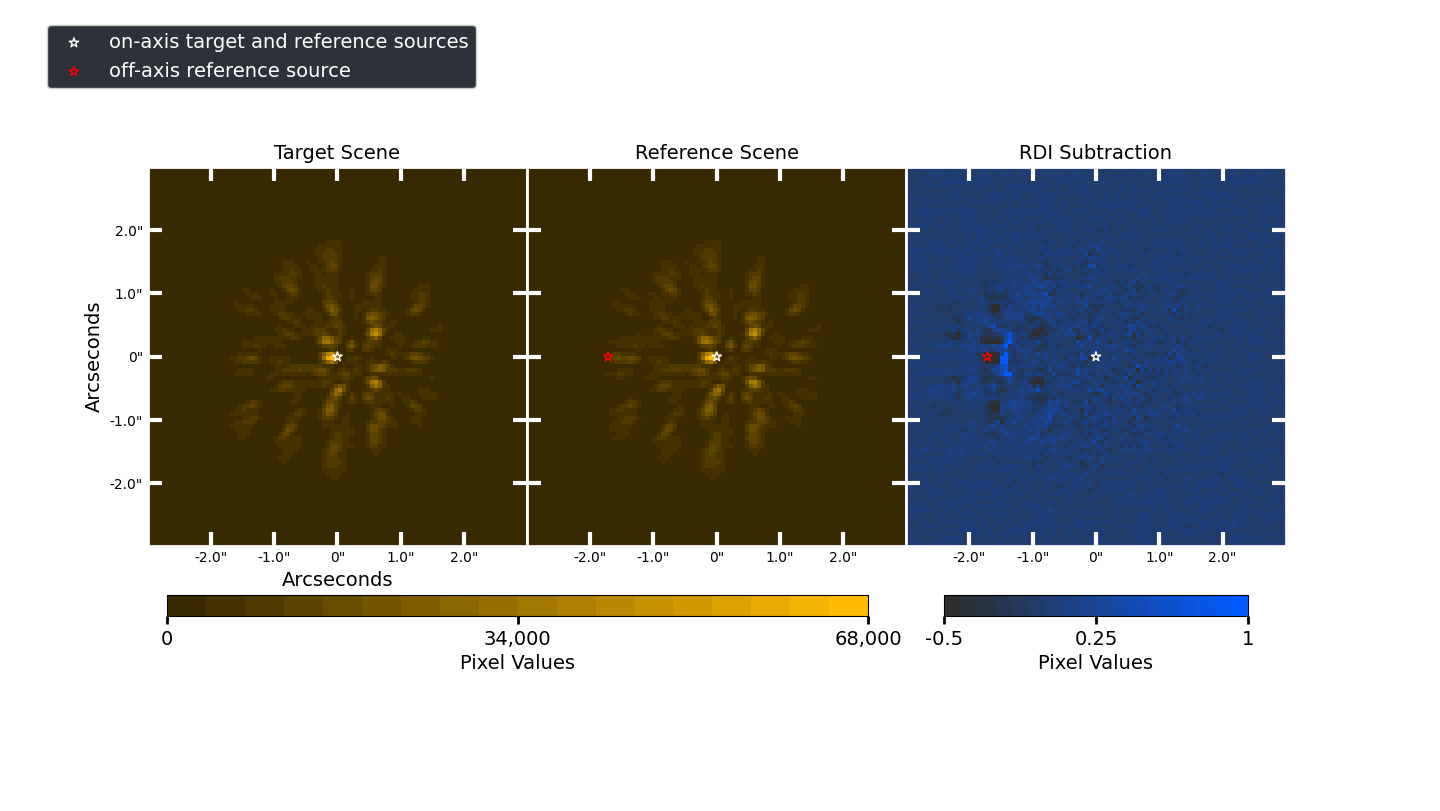}
\centering
\caption{Example target scene, reference scene, and RDI subtraction for a simulation containing an off-axis reference source. \textbf{\textit{Left}}: The .fits file frame of the target scene containing only an on-axis instance of the target star. \textbf{\textit{Center}}: The corresponding reference scene containing both an on-axis instance of the reference star and a binary companion, located 1.5" away from the center of the on-axis source with a relative brightness of $10^{-3}$ that of the on-axis source. \textbf{\textit{Right}}: The RDI subtraction of the two previous frames. The on-axis source is subtracted successfully with very little residuals. The binary companion has some residual flux leftover, indicated by the spattered high valued pixels, indicated with increasingly blue hue assignment. The on-axis sources' positions are indicated by white stars, and the binary companion position is indicated by a red star.} \label{tar-ref-rdi}
\end{figure}

The following is a walk-through for JWST simulations made with \codeword{PanCAKE}, with explanations and justifications of variables as they occur.
\begin{verbatim}
    [1] import pancake
\end{verbatim}
First, import the \codeword{PanCAKE} package, and internally have a proper Anaconda (\cite{anaconda}) environment as needed for \codeword{PanCAKE}'s various package dependencies.
\begin{verbatim}
    [2] target = pancake.scene.Scene('Target')
    [3] target.add_source('HIP 65426', kind='simbad') #inserting on-axis host
\end{verbatim}
This paper explores effects of binaries on RDI subtraction, so we create the first component as necessary for RDI subtraction, a target scene. The target scene contains only one source, HIP 65426. HIP 65426 is used for all simulations in this paper as it was one of the first objects observed with NIRCam, through an Early Release Science Program for direct imaging with JWST (\cite{2023ApJ...951L..20C}). We note that we do not seek to study the effects of binary references on the HIP 65426 system specifically, and we simply use HIP 65426 as a nominal case study example. Additionally, we do not include the companion HIP 65426b, even though it was directly imaged with JWST NIRCam (\cite{2023ApJ...951L..20C}), so that we are sure to isolate the impact of the binary/off-axis reference source. 
\begin{verbatim}
    [3] reference = pancake.scene.Scene('Reference')
    [4] reference.add_source('HIP 65426', kind='simbad') 
    [5] reference.add_source('Companion', kind='grid', r=r_value, theta=theta_value, 
    norm_val=magnitude, spt='a2v', norm_unit='vegamag', norm_bandpass='2mass_ks')
\end{verbatim}
In reality, RDI subtraction utilizes a reference source that is a separate object from the on-axis source. A real reference source is variant from the on-axis in some capacity, whether it be magnitude or spectral type. Focusing on the effect of the off-axis reference source only, we make the on-axis reference source the same object as the on-axis target source, HIP 65426. This allows us to maximize the efficiency of on-axis PSF subtraction, minimizing perceived `degrading contrast' effects caused by using a different on-axis reference source, and focus on the effects of an additional binary companion. The binary companion, labelled `companion' in the lines preceding, is specified using \codeword{PanCAKE}'s grid method; in this case, we select an A2V type star, but vary other properties such as its position and magnitude. The \codeword{r} parameter allows the user to set the location of the binary companion relative to the on-axis host. The \codeword{theta} is the position angle of the binary companion around the on-axis source. 
\begin{verbatim}
    [6] seq = pancake.sequence.Sequence()
    [7] seq.add_observation(target, exposures=[('F444W', 'DEEP8', 18, 5)], 
    nircam_mask='MASK335R', rolls=[0])
    [8] seq.add_observation(reference, exposures=[('F444W', 'DEEP8', 18, 5)], 
    nircam_mask='MASK335R', scale_exposures=target)
\end{verbatim}
Now that we have established our target and reference scenes, we create and perform observations of our scenes. First observing the target, we use optimal readout parameters provided by \codeword{PanCAKE} for an exposure duration of 30 minutes. Our filter choice is NIRCam filter `F444W,' one of the most heavily used filters for NIRCam coronagraphy; it has been adopted for multiple NIRCam surveys/observations for low mass planets (e.g. \cite{2024jwst.prop.6012M,2024jwst.prop.6122B,2023jwst.prop.4050C,2024jwst.prop.5835C,2023jwst.prop.3989H}). For identical reasons, we select the `MASK335R' coronagraphic mask. Simulations using other coronagraphs and filters are out of this work's scope and should be considered in the future. The exposures are repeated for the reference scene containing the on-axis source and binary companion.
\begin{verbatim}
    [9] results = seq.run(save_file=custom_filename, ta_error='none')
    [10] pancake.analysis.contrast_curve(results, target='Target', references='Reference', 
    subtraction='RDI', save_prefix=save_prefix, klip_subsections=10, klip_annuli=10, 
    sub_only=False, regis_err='zero')
\end{verbatim}
The \codeword{ta_error} variable is for target acquisitions. Real JWST target acquisition is imperfect due to slew inaccuracies, and instrument/optical misalignment (\cite{2022SPIE12180E..3QG}, \cite{nelan_2005}). By setting \codeword{ta_error} to `none' \codeword{PanCAKE} eliminates target acquisition error, diminishing the unwanted contribution to our uncertainty factor, ensuring our simulations are as similar as possible. It is in the \codeword{contrast_curve} line where we instruct \codeword{PanCAKE} to perform an RDI subtraction, and set \codeword{pyKLIP}-specific parameters (\cite{2015ascl.soft06001W}), which influence this subtraction. The \codeword{sub_only} parameter turns off contrast curve calculation, allowing only RDI PSF subtraction to occur. It was kept toggled off for a majority of simulations to save runtime. The error introduced when registering unsubtracted images to a common center is \codeword{regis_error}, which accounts for an imperfect estimate of the stellar position across each image. For our simulations we treat it similarly to the \codeword{regis_err} and set it to zero to prevent it from impacting any of our results. 
The three main results from all \codeword{PanCAKE} simulations utilized in this paper are the target and reference simulated observations along with the RDI subtraction of the two observations. Figure \ref{tar-ref-rdi} features these three \codeword{PanCAKE} products for an example off-axis scenario.
\FloatBarrier
\subsection{Contrast Curves} \label{sec:2.5}
\begin{figure}[ht!] 
\includegraphics[width=\textwidth]{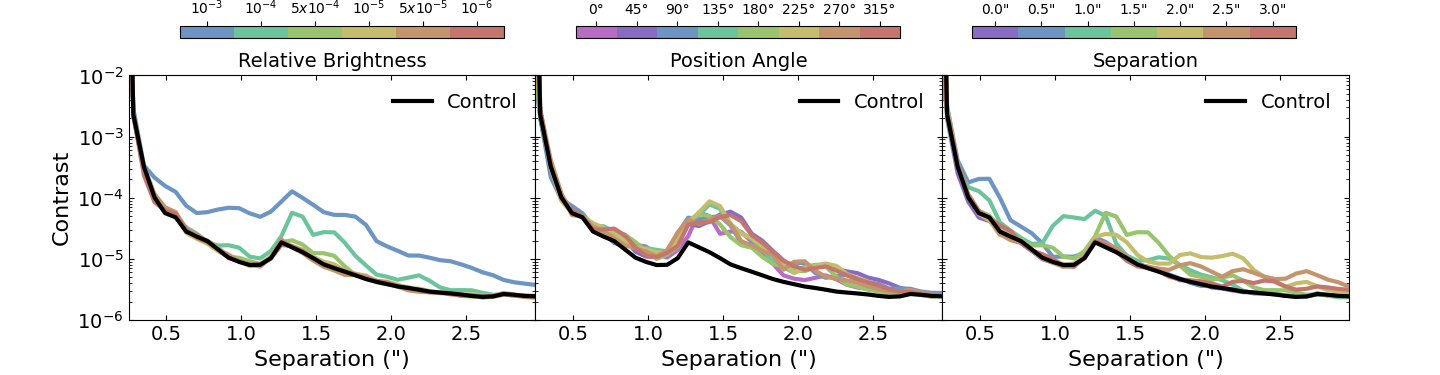}
\centering
\caption{\textbf{\textit{Left}}: Contains contrast curves of binary references as a function of relative brightness, ranging from $10^{-3}$ to $10^{-6}$. The separation of the binary companions in this sub figure are all $1.5"$ from the on-axis source. There are six steps of color assignments correlating to each relative brightness plotted. \textbf{\textit{Center}}: Contains contrast curves of binary references as a function of position angle, ranging from 0° to 315° in steps of 45°. The separation of the binary companions in this sub figure are all $1.5"$ from the on-axis source. The relative brightness of binary companions in this plot are $10^{-4}$ that of the on-axis source.
\textbf{\textit{Right}}: Displays a contrast curve plot showcasing binaries as a function of separation, starting from $0.0"$ to $3.0"$ in steps of $0.5"$. The relative brightness of binary companions in this plot are $10^{-4}$ that of the on-axis source.} \label{contrast-curves}
\end{figure}

To quantitatively represent the degradation of contrast caused in RDI subtraction by the presence of an off-axis reference source, we make use of \codeword{PanCAKE}'s automatically generated \codeword{pyKLIP} based contrast curves. Found in Figure \ref{contrast-curves} are several sets of curves, organized by the relevant binary characteristic being evaluated. The contrast curves are setup using custom script \codeword{custom-curve-plots.py}, which plots multiple contrast curves at once, differentiating each line by use of color assignments. The `control' curve is the average of ten instances of a control \codeword{PanCAKE} scenario containing no additional sources outside of an on-axis target and reference source.
\FloatBarrier

\vspace{3mm}

It is evident in Figure \ref{contrast-curves} that some regions of the image exhibit more severe contrast losses than others, and the binary companion does not impact our RDI subtraction evenly. To showcase the localized effects of binary companions on JWST NIRCam RDI subtraction in an efficient format, we must consider an alternative route to analyzing our RDI subtractions. The following subsections describe a more involved approach to exploring the effects of binary references. 

\subsection{Localized Standard Deviation Calculations \& Contrast} \label{sec:03}
Traditionally contrast is informed by means of a standard deviation within an annulus, centered on the on-axis source or target star. This methodology is limited in the sense that it assumes the stellar noise regime to be relatively static as a function of azimuth. Residual flux varies significantly as a function of azimuth due to the binary companion, and so a more localized approach is needed to analyze our simulations.

For a given pixel, we compute the local standard deviation by considering all surrounding/neighbouring pixels within a circle with radius equal to the spatial resolution, $\lambda/D$. We then compute this $\lambda/D$ spaced standard deviation on a pixel-by-pixel basis across a given RDI subtracted image. This allows us to explore how the noise regime changes across different locations, for example, extremely close to the binary companion versus far away. Here $\lambda$ is the observed wavelength and $D$ is our simulated telescope diameter, an approximated size of 5.2~m due to the obscuring effects of the NIRCam Lyot stop. Lyot stops are substrate layers with holes shaped to correspond with the observations applied occulter. Since light does not transmit anywhere but through the occulter-shaped hole, Lyot stops effectively reduce the aperture of an instrument when used. While different sizes for the circle radius could be chosen, we do not perform such a comparison within this work. In general, smaller radii will be more sensitive to local variations in the noise properties, whereas larger radii act to wash-out those same variations. For this work we have made the assumption that $\lambda/D$ is an appropriate middle-ground, as it corresponds to the scale of an astrophysical feature that one would endeavor to detect.

\begin{figure}[ht!]
\raisebox{8mm}{\includegraphics[width=49mm]{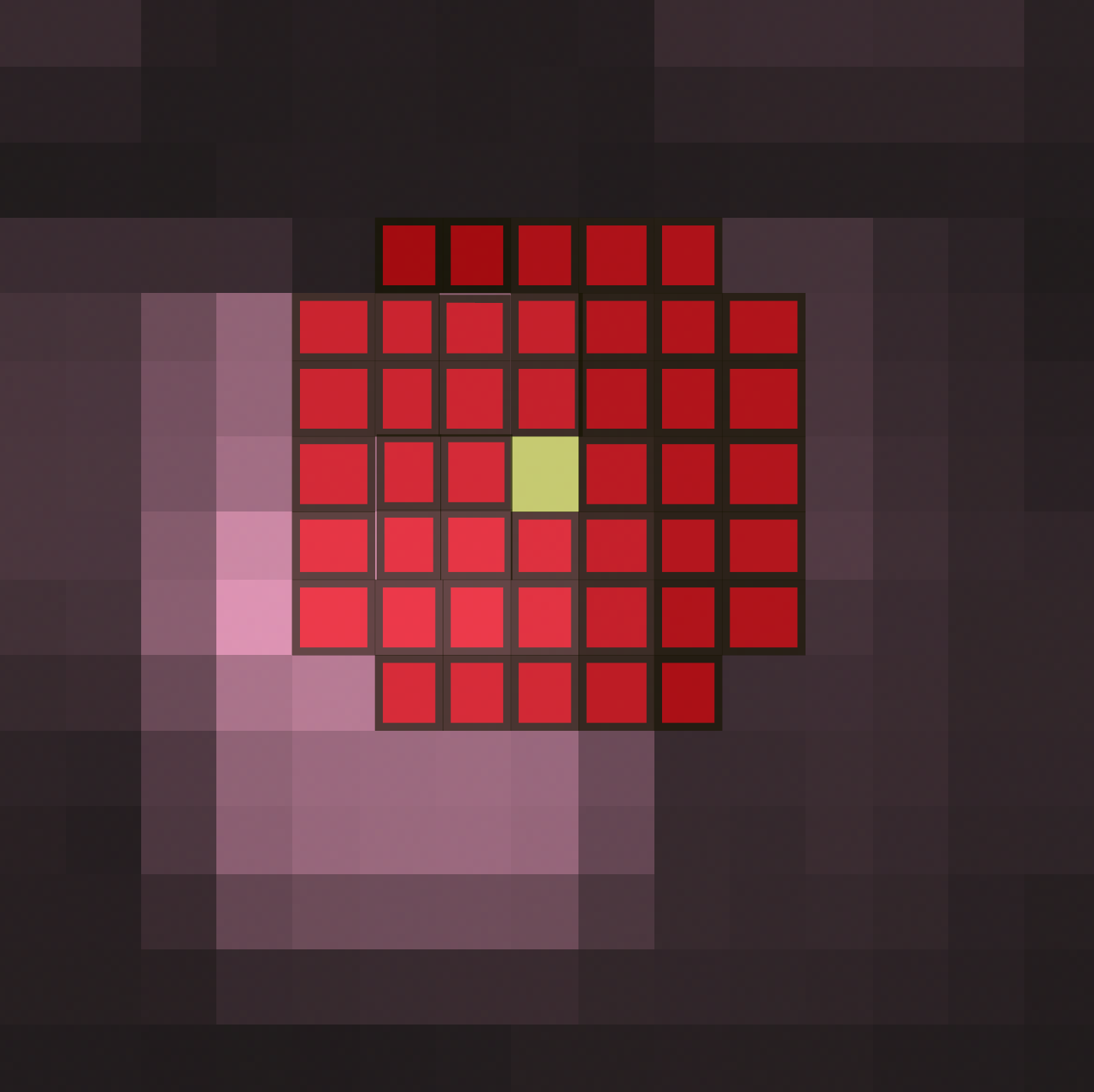}}
\includegraphics[width=9cm]{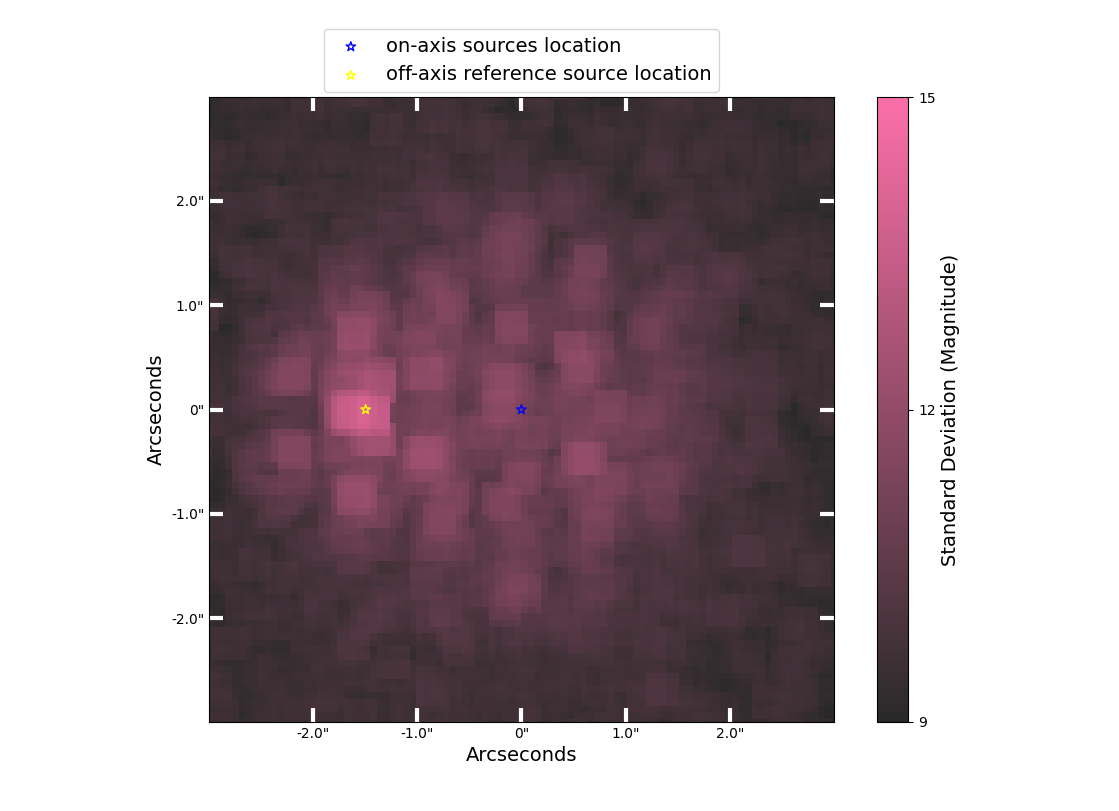}
\centering
\caption{\textbf{\textit{Left}}: A diagram representing the 2D computation of standard deviations for the paper. A pixel of interest located in the center of a binary companion, is indicated in yellow. The surrounding red ring-shaped mask is made of valid pixels within a $\lambda$/D distance away that are included in the calculation of the yellow pixel-of-interest's standard deviation. \textbf{\textit{Right}}: The 5$\sigma$ standard deviation calculation for an RDI subtraction containing a binary companion at a 1.5" separation from the center of the on-axis sources, with a relative brightness of $10^{-4}$ that of HIP 65426.} \label{standard-deviation-science}
\end{figure}

After finding the standard deviation, we then perform a few array operations using a custom created script. This program is found on the paper's GitHub page \cite{klaus-stephenson_2024}. First, multiplication of the standard deviation array to achieve a 5$\sigma$ units. We then divide the array values by the peak off-axis flux found from our original host star to estimate a 5$\sigma$ contrast. The peak off-axis flux is taken from \codeword{PanCAKE}'s \codeword{analysis.py} file, in line $323$ from within the \codeword{extract_simulated_images} function. Since contrast represents the flux ratio between a star and a feasibly detectable object, we then use
\begin{equation}
    m_1=-2.5log(\frac{f_1}{f_2})+m_2
\end{equation}
to convert the 5$\sigma$ contrast into units of apparent magnitude sensitivity. Expanding upon the example scenario used in the previous section, the 5$\sigma$ standard deviation computed for the example's RDI subtraction is also shown in Figure \ref{standard-deviation-science}.

\FloatBarrier

\subsection{Magnitude Sensitivity Loss} \label{sec:04}
To quantitatively represent the change caused by binary companions, we consider the sensitivity loss, in units of magnitude, caused by a binary companion. The sensitivity loss compares a simulation of an RDI subtraction containing a binary reference to a `perfect' control scenario of a single-PSF reference; demonstrating how much worse a subtraction containing a binary reference is compared to a scenario with no binary companion at all. 

\begin{figure}[ht!]
\includegraphics[width=15cm]{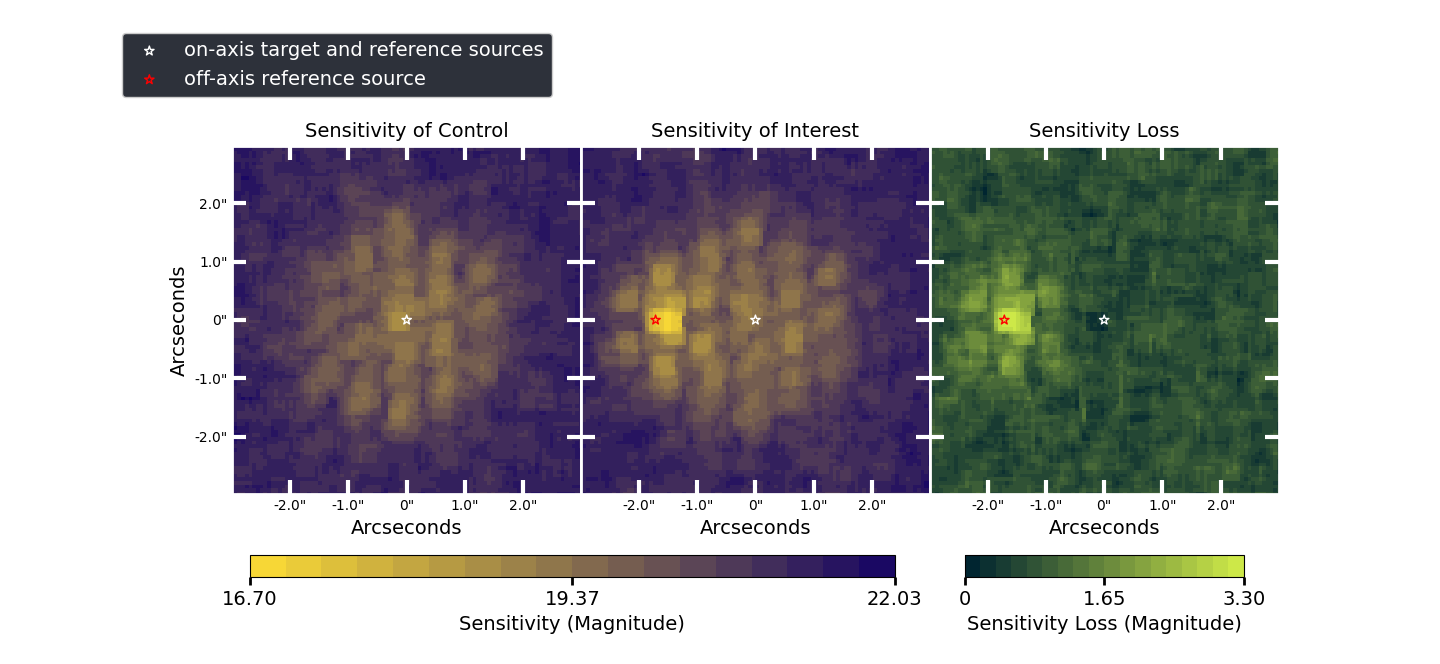}
\centering
\caption{\textbf{\textit{Left}}: The 5$\sigma$ standard deviation of a control scenario containing no binary reference companion. \textbf{\textit{Middle}}:, the 5$\sigma$ standard deviation of our example scenario containing a binary companion at 1.5" away from the on-axis host with a relative brightness of $10^{-4}$. \textbf{\textit{Right}}:~The magnitude loss caused by the binary companion. The rightmost plot is a subtraction of the middle plot from the leftmost plot.} \label{std-std-msl}
\end{figure}

A common control 5$\sigma$ magnitude sensitivity was used to calculate all sensitivity losses in this paper. The sensitivity loss is the 5$\sigma$ sensitivity of interest subtracted from the 5$\sigma$ sensitivity of a control scenario. The sensitivity loss calculated for our example scenario is found in Figure \ref{std-std-msl}.
\FloatBarrier
Once the magnitude loss is found for an RDI observation, we consider the total average magnitude sensitivity loss, or `total average loss,' and the local average magnitude sensitivity loss, or `local average loss'. The total average loss informs users of the average magnitude sensitivity loss within the 3"x3" simulated region due to the presence of a binary companion. The local average loss informs users of the average magnitude sensitivity loss within one 1" of the center of a binary companion. 1" was chosen to encapsulate the unique multi-lobed JWST NIRCam coronagraphic PSF while staying local to the most extreme binary effects.

The total average loss in our example science scenario is $0.10$, and the local average loss is $1.05$. The former value enlightens us on how an off-axis reference PSF impacts the final sensitivity both generally/across the entire image, whereas the latter value is telling of a more extreme sensitivity loss induced locally to the binary companion. While these values on their own are informative, they are but one example of the magnitude sensitivity loss caused by one specific set of binary parameters. To obtain a more holistic view of the effects of binary companions on JWST NIRCam RDI coronagraphy, we must consider and evaluate the magnitude sensitivity loss caused by other instances of binary characteristics.

\subsection{Uncertainty calculation} \label{sec:4.5}
The largest factor of uncertainty in our simulation work stems from the variance between \codeword{PanCAKE} simulations. There is variability between simulation runs even when limiting factors that would otherwise contribute further to uncertainty, such as \codeword{ta_err} and \codeword{regis_err}. To calculate the uncertainty caused by the inherent source code of \codeword{PanCAKE}, we first simulated 20 different instances of our staple example science scenario containing a binary companion at 1.5" separation with a relative brightness of $10^{-4}$ that of the on-axis source. We compared each sensitivity loss value, both on a local and total scale, produced by each instance of the same binary simulation. We then did the reverse, re-calculating the sensitivity loss on both local and total scales for 20 different instances of the control scenario, keeping the staple science scenario static. 

The standard deviation of sensitivity loss values caused by changing the staple science scenario on a total scale is $0.0070$~mag, and on a local scale is $0.0155$~mag. Getting different results when we change the control scenario, we found the standard deviation of sensitivity loss values on a total scale is $0.0316$~mag, and on a local scale is $0.0552$~mag. 

This difference between the standard deviation for sensitivity loss caused by changing instances of the control versus the science scenario is not well understood. We suspect that adding the binary companion changes the nature of \codeword{PanCAKE} subtraction, restricting the variations between binary-containing simulations. Further work is needed to explore these variations in more detail. 

For this paper's scope we consider the uncertainty to be $0.057$ locally, and $0.032$ on a total scale, both values in units of magnitude. In cases where a sensitivity loss is less than zero and the absolute value is within the respective uncertainty parameter we manually redefine it as $0$, as adding a binary companion should not improve sensitivity. The largest negative value within our simulations for a local loss value was $-0.059$ which is within 1$\sigma$ of the $0.057$ local uncertainty parameter. The largest negative value within our simulations for a total loss value was $-0.01$ which is within the $0.032$ total uncertainty parameter.

\section{Results} \label{discussion}
Moving away from our case science example, we apply the processes described prior to several ranges of binary characteristics. The characteristics evaluated are magnitude from $10^{-3}$ to $10^{-6}$, separation starting from $0.0"$ (or $0.5"$) extending to $2.5"$ (or $3.0"$), and lastly position angle relative to the on-axis source from 0° to 315°. Extensive `heatmap' plots showcasing the total and local average magnitude sensitivity losses can be found here. The effects of binary companion magnitude as a function of separation from on-axis sources can be found in Figure \ref{heatmap-01}, and the effects of binary companion position angle as a function of separation is found in Figures \ref{heatmap-02}, \ref{heatmap-03}, and \ref{heatmap-04}. 
\subsection{Varying binary companion Magnitude}
\begin{figure}[ht!] 
\includegraphics[width=14cm]{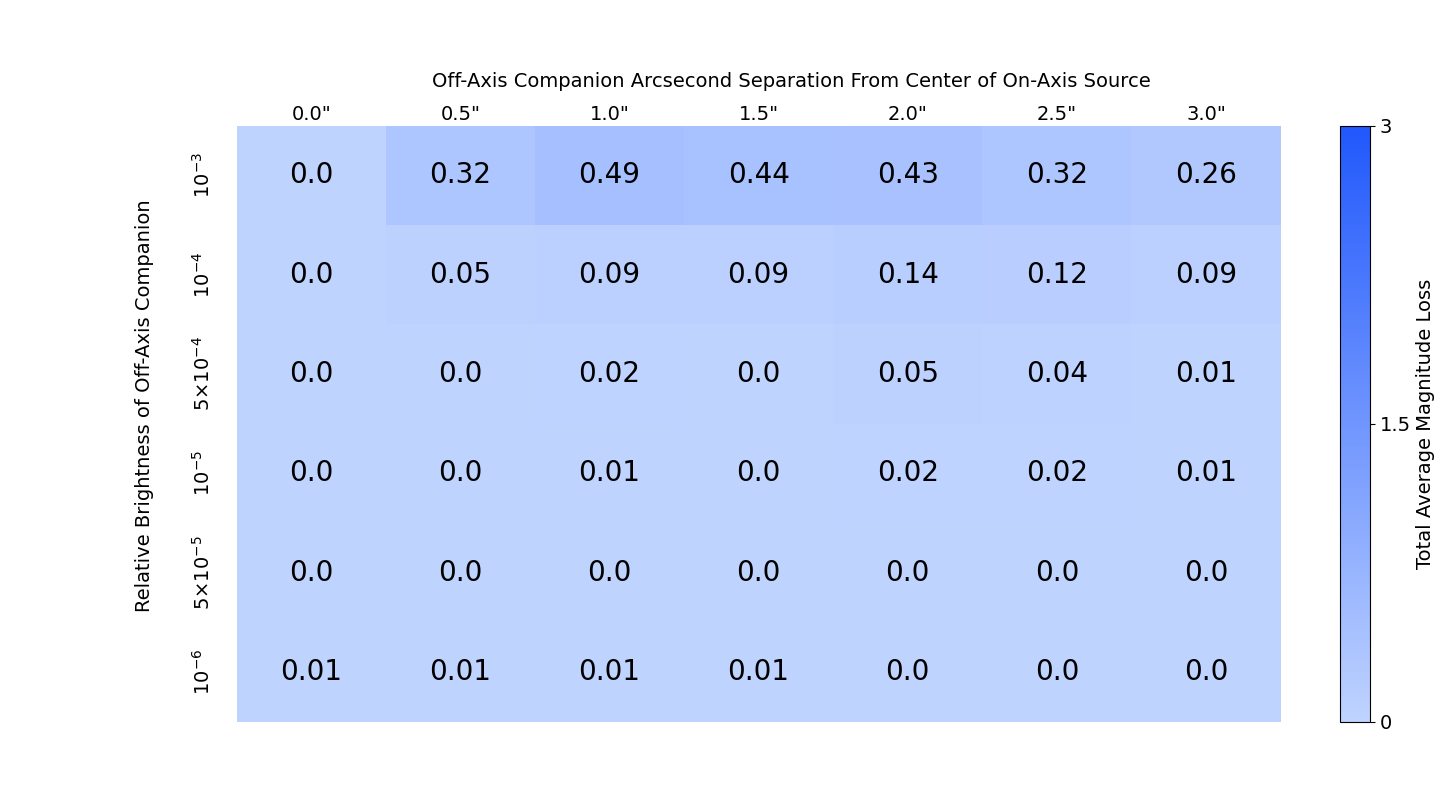}
\includegraphics[width=14cm]{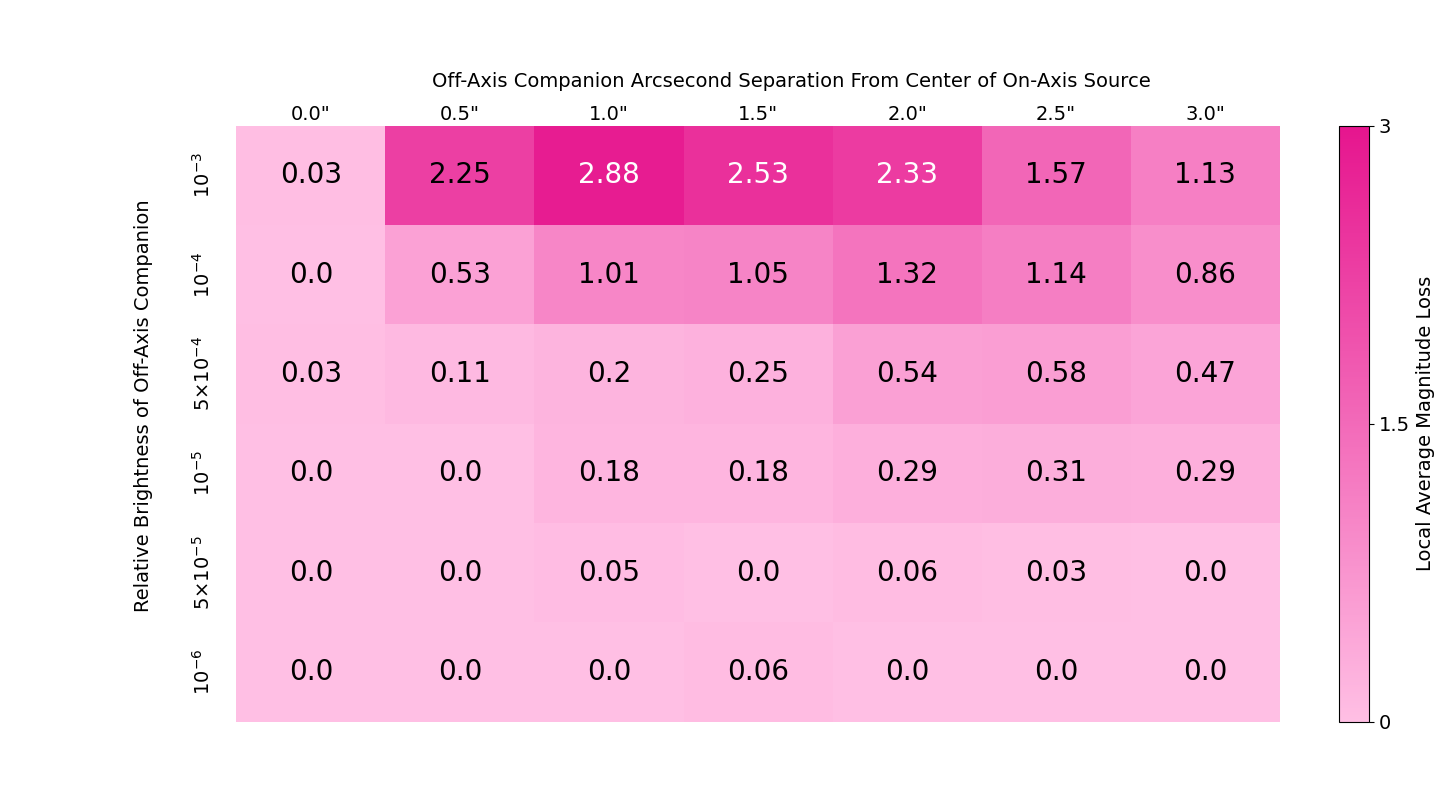}
\centering
\caption{Magnitude loss for varying binary companion magnitudes as a function of separation. The blue colored plot is total average magnitude sensitivity loss, the average loss over an entire image. The pink colored plot is local average magnitude sensitivity loss, or the average loss within 1" of the binary companion. All values for the local plot have an uncertainty of $\pm 0.059$; all values for the total average loss plot have an uncertainty of $\pm 0.032$. } \label{heatmap-01}
\end{figure}

To analyze the effects of binary companions as a function of brightness, we evaluated binaries with separations ranging from $0"$ all the way to our limiting FOV of 3". We include $0"$ separated binary companions to represent binary systems whose companion is separated by some minuscule amount not resolvable by JWST. Referencing Figure \ref{heatmap-01}, the stark difference in sensitivity loss caused on a total scale compared to loss on a local scale is clear. This is accentuated for binaries with high relative brightness, particularly those within the $10^{-3}$ to $10^{-4}$ regime. The worst sensitivity loss induced by changing a binary companion's relative brightness is $2.88$~mag, and is caused by a binary whose companion is $1.0"$ away from the center of the binary companion, with a relative brightness of $10^{-3}$. The corresponding total loss is $0.49$~mag, and is similarly the worst sensitivity loss for this set of simulations. Similar sensitivity losses, on both total and local scales, are exhibited by binary companions within $1.0"$ to $2.5"$. This makes sense as these binary companions are fully resolvable within the simulation FOV, not occulted by the coronagraph centered over the on-axis source, nor partially to fully cropped by the image border. NIRCam's real FOV is 10"$\times$10" but for our simulations we adopt a much smaller FOV since \codeword{PanCAKE} is limited to 3"$\times$3". 

It appears that binaries with a relative brightness of $5\mathrm{x}10^{-4}$ or less cause little to no degradation in sensitivity, on a total scale. We begin seeing local average sensitivity loss values of $0.0$ magnitude starting at $10^{-5}$; an exception being binaries whose companion is at a $0"$ separation from the on-axis source, which makes sense as those binaries are essentially a \codeword{PanCAKE} simulation with an on-axis reference source slightly brighter than the on-axis target source. A closely separated binary could be recommended for observers imaging widely separated targets; by taking advantage of the binary companion being partially occulted by a coronagraph one would minimize the binary companion's impact on sensitivity loss. However, this method would not be advisable for observers imaging planets/objects close to the coronagraph as you would be magnifying the binary companion's influence on the science object of interest, regardless of the binary companion's partially occulted state.

Relating back to Figure \ref{contrast-curves}, it should be noted that the NIRCam instrument's contrast capability already diminishes naturally at shorter separations. Meaning that although values of magnitude loss may be smaller at 0.5" compared to 1" in figure \ref{heatmap-01}, the actual magnitude sensitivity is likely to be worse; this should be considered when opting to use a binary reference by those seeking to detect faint companions at short angular separations.
\FloatBarrier
\subsection{Varying Off-Axis position angle}
\begin{figure}[ht!]
\includegraphics[width=14cm]{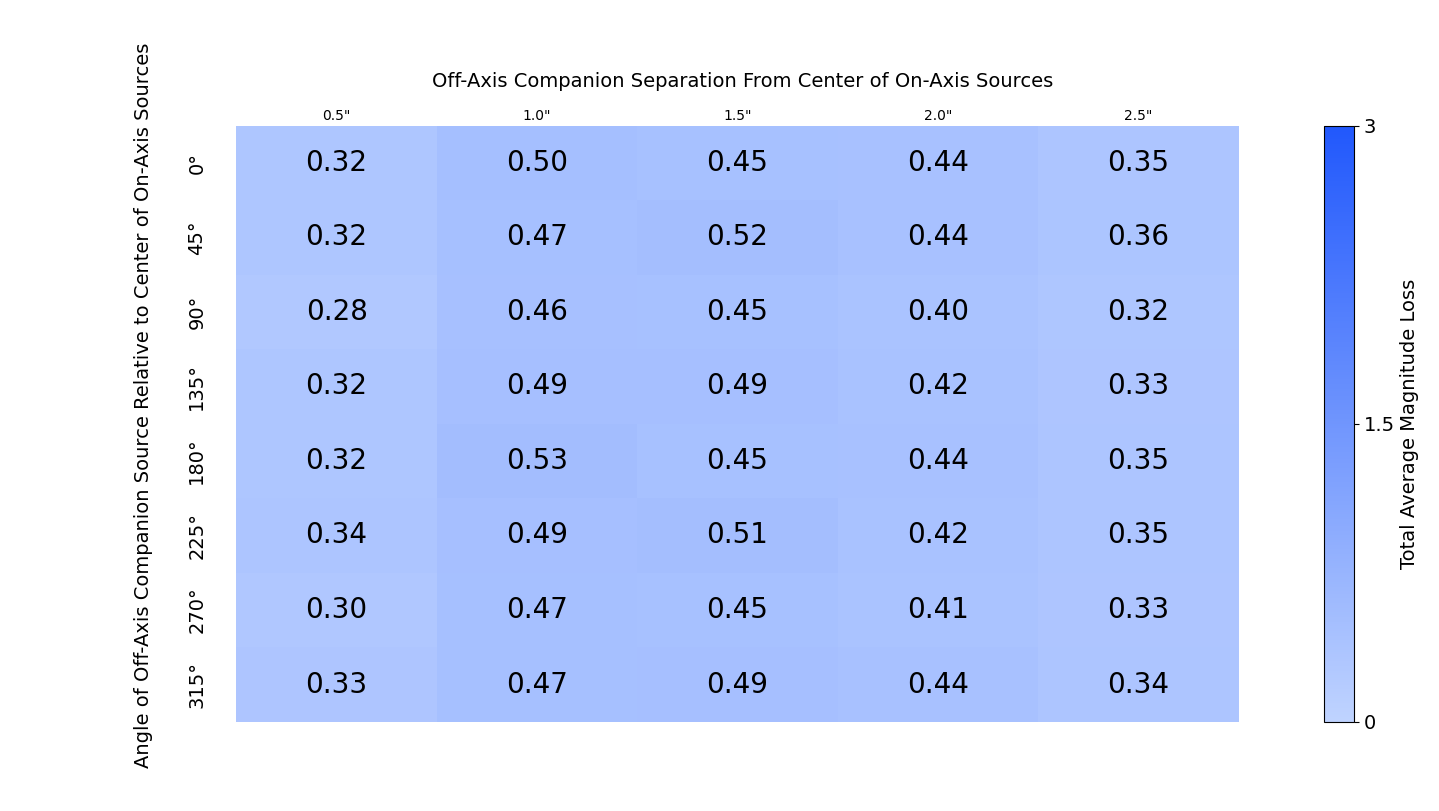}
\includegraphics[width=14cm]{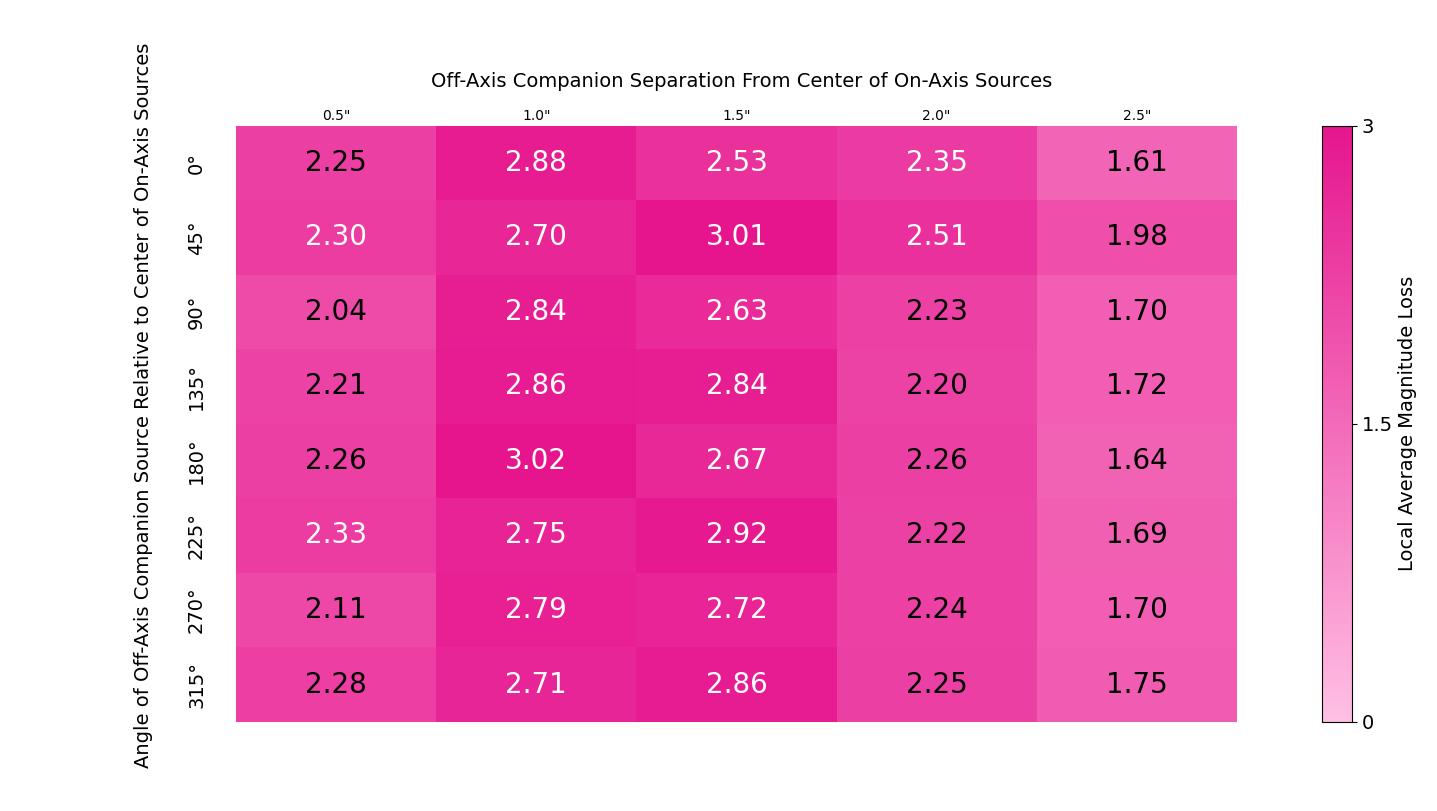}
\centering
\caption{Magnitude loss for varying binary companion position angles as a function of separation, with the binary companion relative brightness being $10^{-4}$ that of the on-axis source. The blue colored plot is total average magnitude sensitivity loss, the average loss over an entire image. The pink colored plot is local average magnitude sensitivity loss, or the average loss within 1" of the binary companion. All values for the local plot have an uncertainty of $\pm 0.059$; all values for the total average loss plot have an uncertainty of $\pm 0.032$. } \label{heatmap-02}
\end{figure}

\begin{figure}[ht!] 
\includegraphics[width=14cm]{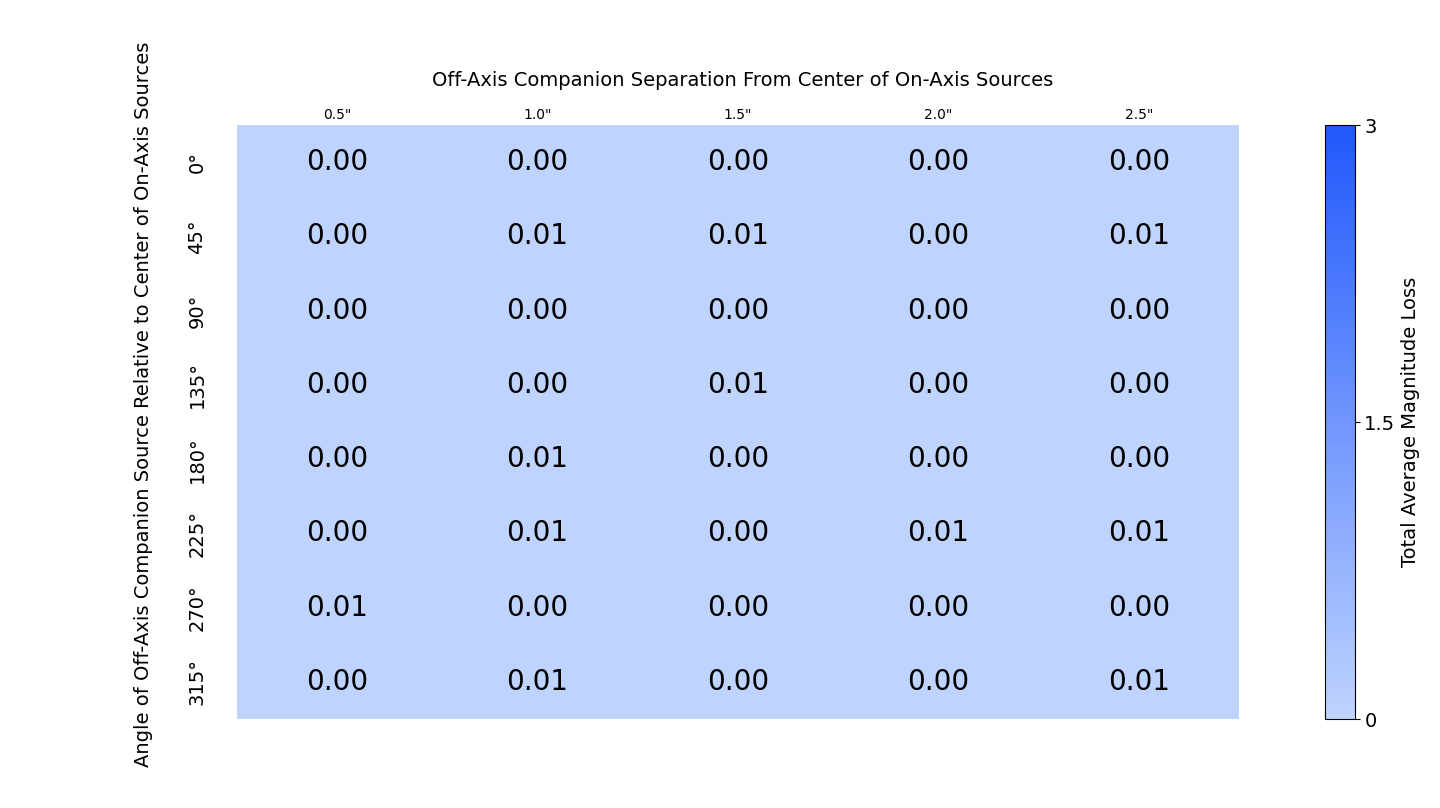}
\includegraphics[width=14cm]{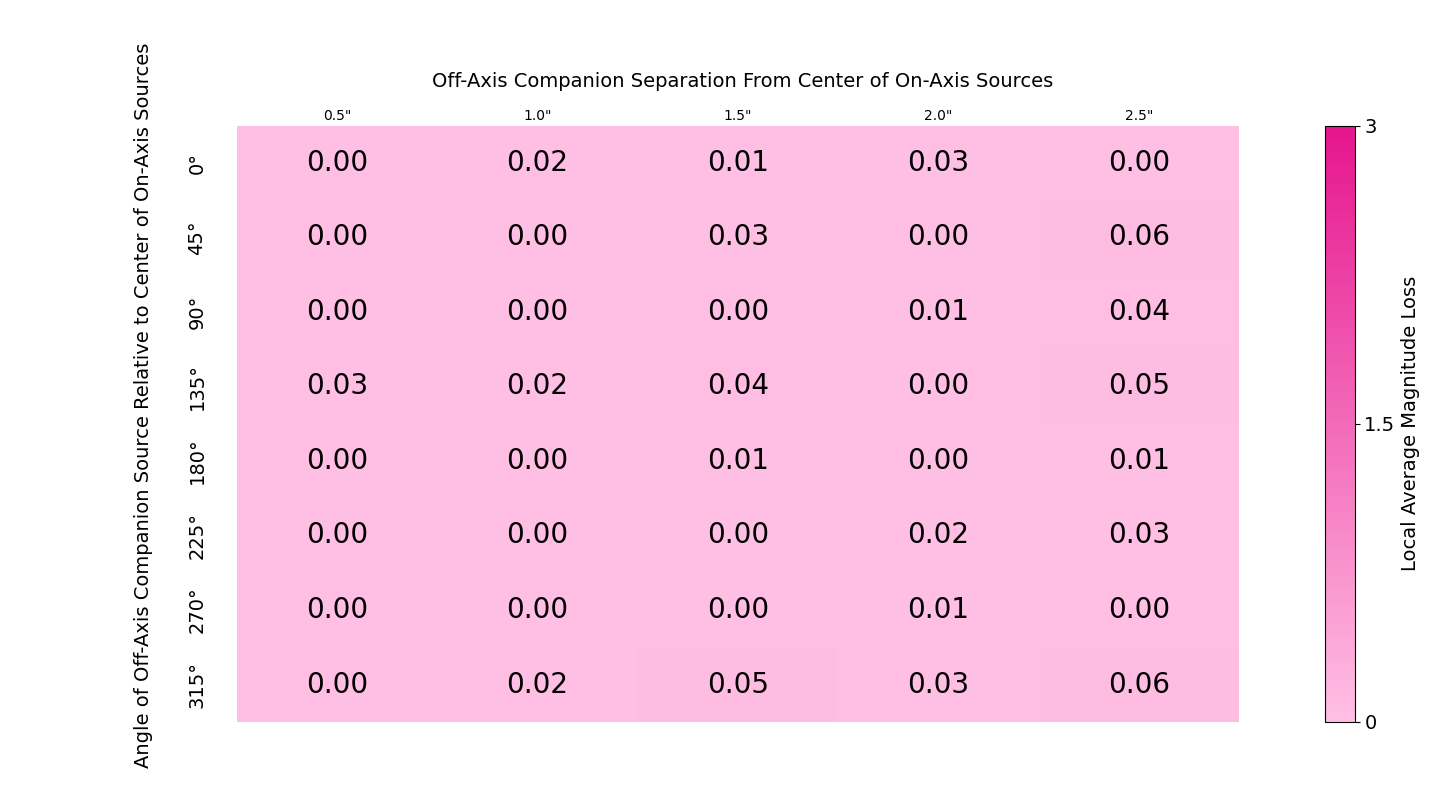}
\centering
\caption{Magnitude loss for varying binary companion position angles as a function of separation, with the binary companion relative brightness being $10^{-6}$ that of the on-axis source. The blue colored plot is total average magnitude sensitivity loss, the average loss over an entire image. The pink colored plot is local average magnitude sensitivity loss, or the average loss within 1" of the binary companion. All values for the local plot have an uncertainty of $\pm 0.059$; all values for the total average loss plot have an uncertainty of $\pm 0.032$. } \label{heatmap-03}
\end{figure}

In Figure \ref{heatmap-02}, we show the average magnitude sensitivity loss caused as a function of position angle and separation. All of the binaries in this set of simulations have a relative brightness of $10^{-4}$ that of the on-axis source. The highest average sensitivity loss occurs at $1.0"$ at a position angle of $180^\circ$, with a total average magnitude loss of $0.53$~mag, and local average sensitivity loss of $3.02$~mag. Magnitude sensitivity loss does not dip below $0.28$ across neither the total nor local average plots. The least degrading binary out of this set of simulations occurs at $0.5"$ with a position angle of $90^\circ$, with a total magnitude sensitivity loss of $0.28$~mag and a local loss of $2.043$~mag. Referencing \ref{heatmap-02}, we can see in this `heatmap' and the ones following, that JWST's uniquely spatially structured PSF impacts our sensitivity loss results depending on whether or not the binary companion falls off or onto a PSF lobe.
\FloatBarrier
Our last `heatmap' is figure \ref{heatmap-03}, featuring varying position angles/separations of a binary companion with a relative brightness of $10^{-6}$ that of the on-axis source. The most severe total average sensitivity loss in this set of simulations is $0.1$ units of magnitude, occurring at numerous separations and rotations. The most severe local average sensitivity loss occurs twice at $2.5"$ at position angles of 45° and 315°, sensitivity loss being $0.06$. 

While position angle does play a role in magnitude sensitivity loss, causing $0.05-0.3$ magnitude sensitivity variations, its impact is much smaller compared to the influence of separation and relative magnitude.
\FloatBarrier

\vspace{3mm}
In summary, binaries inducing the least sensitivity loss occur at the smallest separations (leftmost columns in the figures), tending to be $0.0"$ or $0.5"$ depending on whether one is looking at the varying magnitude `heatmap' or the varying position angles `heatmaps.' The second least degrading binary separation is $3.0"$, where the binary companion is cropped/limited by the outer border of the simulation FOV \footnote{$3.0"$ is only featured in the first `heatmap' due to position angle simulations at this separation generating errors.}. 

Generally, brighter binary companions result in more extreme sensitivity losses. The highest overall total and local average magnitude sensitivity loss were by a binary companion with a relative brightness of $10^{-3}$ compared to the on-axis source, those values being $0.53$~mag and $3.02$~mag respectively. Observations of $0.0$ sensitivity both a total and local scale begin to occur at $10^{-5}$ relative brightness. The $10^{-5}$ relative brightness `heatmap' as well as a $10^{-4}$ one can be found in Appendix \ref{bonus}.

\section{Discussion \& Conclusion} \label{conclusion}
JWST's limited nominal lifetime, originally projected to be 5-10 years, is now projected to be closer to 20~years due to leftover launch fuel (\cite{fox_2021}). However, many factors could contribute to the failure of major component(s) at any given time, and optimizing proposal efficiency is critical for community members seeking to observe with JWST. JWST is in high demand as it provides unprecedented contrast for the faintest observable objects yet and it is crucial for members of the astronomy community to put JWST to use as best they can, while they can. Those considering RDI imaging for their proposals must concern themselves with the practicality of using single-source PSFs when the opportunity or need to use a moderate contrast binary presents itself. The results of this paper, the effects of binary references on JWST NIRCam RDI subtraction, are meant to serve as a reference index for observers planning JWST RDI proposals. The `heatmap' plots are an opportunity to inform those of the quantifiable impact of binary references in their JWST NIRCam direct imaging campaign(s). Our general recommendations are as follows:

\textbf{\textit{Considering Magnitude}} -- Utilize the dimmest binary accessible. binary companions whose relative brightness is between $10^{-3}$ and $10^{-4}$ that of their on-axis counterpart produce the highest sensitivity losses. However, it should be noted that a relative brightness of $10^{-3}$ corresponds to the brightest binary companion explored in this work. $10^{-3}$ relative brightness binaries produce magnitude sensitivity losses between $0.28$ and $0.53$ on a total average scale, and $1.64$ to $3.02$ on a local (within $1"$ of binary companion) average scale. $10^{-4}$ relative brightness binaries produce magnitude sensitivity losses between $0.01$ and $0.17$ on a total average scale, and $0.37$ to $1.45$ on a local average scale. Binaries dimmer than $10^{-4}$ may be opportunistic to observers as there are several separations at this relative brightness where there is $0.00$ magnitude sensitivity loss on a total scale. $0.00$ average magnitude sensitivity loss on a local scale does not begin to occur until the binary companion has a relative brightness of $10^{-5}$ the on-axis source.

\textbf{\textit{Considering Separation}} -- This trade space should be considered on a case-by-case basis and therefore we make no 'general' recommendations. By using the closest separated binaries, assuming science goals allow for it, one can capitalize on binary companion cropping performed by the coronagraph. However, it is at locations closest to the coronagraph that observers my be most interested in imaging an exoplanet. It is also at close separations where NIRCam sensitivity decreases, meaning effects of binary companions may be more impactful here despite smaller sensitivity loss. Exceedingly separated binaries that neared our simulation FOV achieved similarly cropped results, and it is also at wider separations that NIRCam is most sensitive. Binary companions between the inner and outer simulation FOV have are fully resolvable, therefore leaving the most amount of residual flux. 

\textbf{\textit{Considering Position Angle}} -- Varying position angle of a binary companion has impact on sensitivity loss, due to the unique JWST lobe-shaped PSF. Change in sensitivity loss induced locally deviates by $0.3-0.5$ units of magnitude for a given separation or relative brightness. 

There are several scenarios where binaries are suitable references, as well as several contexts where using a binary reference is ill advised. Binaries of low relative brightness, $10^{-5}$ to $10^{-6}$, relative to the on-axis PSFs, are capable of inducing $0.00$ average magnitude sensitivity loss on the overall image, making these binaries suitable reference sources for most coronagraphic observations. Binaries of close separations of $\lesssim$ $0.5"$, have the next least amount of impact on RDI subtractions. Binaries of extremely large separations which border, or are cropped by, the effective simulation FOV have similarly less impact compared to companions fully resolved within the observation FOV.

It is generally not advisable to use a binary for RDI subtraction if it is of relative brightness greater than or equal to $10^{-4}$ to the on-axis source. However it's feasible to use brighter binaries for those imaging widely separated targets, located closely to the coronagraph. In all cases is not advisable to use a binary if it will be closely located to a science object/feature of interest, as it's at these locations where sensitivity loss is most dramatic. Those seeking to image known systems with predictable orbits should purposefully binaries only if they are able to reliably time their observations so that the desired imaged object(s) is clearly separated from the binary companion, or at a minimum, carefully consider the negative impacts on their ability to detect and/or characterize that object in the presence of binary contamination. 

Overall, we expect the results searchable in Section \ref{discussion} and appendix \ref{bonus} to serve JWST observers as a guide to the effects of binary companions given various parameters and ultimately leave the work open ended for readers to weigh the various deficits themselves. Work completed in this paper should better inform observers and their decisions throughout the JWST proposal process, and also aid in identifying and analyzing the effects of binary references in their own works. Finally, we note that the insights shown in this work represent a worst-case scenario, where a binary companion is present and must be included in the reference subtraction. Further steps to model and subtract the binary companion prior to the PSF subtraction should be explored, and may significantly reduce potential sensitivity losses. Future works should additionally seek to explore the effects of the spectral type of the off-axis reference PSFs, and the effects of binaries for other JWST instruments (i.e. MIRI).

\acknowledgments 
Thank you to Dr. Rebecca Jensen-Clem for insightful conversations on contrast curves, and to Dr. Rachel Bowens-Ruben for discussions of contrast maps and questions on 0.1 M$_\mathrm{Jup}$ mass imaging campaigns. This work made use of \codeword{numpy}, \codeword{WebbPSF}, \codeword{astropy}, \codeword{scipy}, and \codeword{pyKLIP}. This project was supported by a grant from STScI (JWST-ERS-01386) under NASA contract NAS5-03127. 
\section{References}
\bibliography{report} 

\begin{thebibliography}{10}

\bibitem{2023ASPC..534..799C}
{Currie}, T., {Biller}, B., {Lagrange}, A., {Marois}, C., {Guyon}, O.,
  {Nielsen}, E.~L., {Bonnefoy}, M., and {De Rosa}, R.~J., ``{Direct Imaging and
  Spectroscopy of Extrasolar Planets},'' in [{\em Protostars and Planets
  VII}{\nolinebreak\hspace{0.1em}]},  {Inutsuka}, S., {Aikawa}, Y., {Muto}, T.,
  {Tomida}, K., and {Tamura}, M., eds., {\em Astronomical Society of the
  Pacific Conference Series} {\bf 534},  799 (July 2023).

\bibitem{2016PASP..128j2001B}
{Bowler}, B.~P., ``{Imaging Extrasolar Giant Planets},'' {\em \pasp}~{\bf 128},
   102001 (Oct. 2016).

\bibitem{2014PNAS..11112661M}
{Macintosh}, B., {Graham}, J.~R., {Ingraham}, P., {Konopacky}, Q., {Marois},
  C., {Perrin}, M., {Poyneer}, L., {Bauman}, B., {Barman}, T., {Burrows},
  A.~S., {Cardwell}, A., {Chilcote}, J., {De Rosa}, R.~J., {Dillon}, D.,
  {Doyon}, R., {Dunn}, J., {Erikson}, D., {Fitzgerald}, M.~P., {Gavel}, D.,
  {Goodsell}, S., {Hartung}, M., {Hibon}, P., {Kalas}, P., {Larkin}, J.,
  {Maire}, J., {Marchis}, F., {Marley}, M.~S., {McBride}, J.,
  {Millar-Blanchaer}, M., {Morzinski}, K., {Norton}, A., {Oppenheimer}, B.~R.,
  {Palmer}, D., {Patience}, J., {Pueyo}, L., {Rantakyro}, F., {Sadakuni}, N.,
  {Saddlemyer}, L., {Savransky}, D., {Serio}, A., {Soummer}, R.,
  {Sivaramakrishnan}, A., {Song}, I., {Thomas}, S., {Wallace}, J.~K.,
  {Wiktorowicz}, S., and {Wolff}, S., ``{First light of the Gemini Planet
  Imager},'' {\em Proceedings of the National Academy of Science}~{\bf 111},
  12661--12666 (Sept. 2014).

\bibitem{2017A&A...605L...9C}
{Chauvin}, G., {Desidera}, S., {Lagrange}, A.~M., {Vigan}, A., {Gratton}, R.,
  {Langlois}, M., {Bonnefoy}, M., {Beuzit}, J.~L., {Feldt}, M., {Mouillet}, D.,
  {Meyer}, M., {Cheetham}, A., {Biller}, B., {Boccaletti}, A., {D'Orazi}, V.,
  {Galicher}, R., {Hagelberg}, J., {Maire}, A.~L., {Mesa}, D., {Olofsson}, J.,
  {Samland}, M., {Schmidt}, T.~O.~B., {Sissa}, E., {Bonavita}, M., {Charnay},
  B., {Cudel}, M., {Daemgen}, S., {Delorme}, P., {Janin-Potiron}, P., {Janson},
  M., {Keppler}, M., {Le Coroller}, H., {Ligi}, R., {Marleau}, G.~D.,
  {Messina}, S., {Molli{\`e}re}, P., {Mordasini}, C., {M{\"u}ller}, A.,
  {Peretti}, S., {Perrot}, C., {Rodet}, L., {Rouan}, D., {Zurlo}, A.,
  {Dominik}, C., {Henning}, T., {Menard}, F., {Schmid}, H.~M., {Turatto}, M.,
  {Udry}, S., {Vakili}, F., {Abe}, L., {Antichi}, J., {Baruffolo}, A.,
  {Baudoz}, P., {Baudrand}, J., {Blanchard}, P., {Bazzon}, A., {Buey}, T.,
  {Carbillet}, M., {Carle}, M., {Charton}, J., {Cascone}, E., {Claudi}, R.,
  {Costille}, A., {Deboulbe}, A., {De Caprio}, V., {Dohlen}, K., {Fantinel},
  D., {Feautrier}, P., {Fusco}, T., {Gigan}, P., {Giro}, E., {Gisler}, D.,
  {Gluck}, L., {Hubin}, N., {Hugot}, E., {Jaquet}, M., {Kasper}, M., {Madec},
  F., {Magnard}, Y., {Martinez}, P., {Maurel}, D., {Le Mignant}, D.,
  {M{\"o}ller-Nilsson}, O., {Llored}, M., {Moulin}, T., {Orign{\'e}}, A.,
  {Pavlov}, A., {Perret}, D., {Petit}, C., {Pragt}, J., {Puget}, P., {Rabou},
  P., {Ramos}, J., {Rigal}, R., {Rochat}, S., {Roelfsema}, R., {Rousset}, G.,
  {Roux}, A., {Salasnich}, B., {Sauvage}, J.~F., {Sevin}, A., {Soenke}, C.,
  {Stadler}, E., {Suarez}, M., {Weber}, L., {Wildi}, F., {Antoniucci}, S.,
  {Augereau}, J.~C., {Baudino}, J.~L., {Brandner}, W., {Engler}, N., {Girard},
  J., {Gry}, C., {Kral}, Q., {Kopytova}, T., {Lagadec}, E., {Milli}, J.,
  {Moutou}, C., {Schlieder}, J., {Szul{\'a}gyi}, J., {Thalmann}, C., and
  {Wahhaj}, Z., ``{Discovery of a warm, dusty giant planet around HIP 65426},''
  {\em \aap}~{\bf 605},  L9 (Sept. 2017).

\bibitem{2008Sci...322.1348M}
{Marois}, C., {Macintosh}, B., {Barman}, T., {Zuckerman}, B., {Song}, I.,
  {Patience}, J., {Lafreni{\`e}re}, D., and {Doyon}, R., ``{Direct Imaging of
  Multiple Planets Orbiting the Star HR 8799},'' {\em Science}~{\bf 322},  1348
  (Nov. 2008).

\bibitem{2024AJ....167..182C}
{Cugno}, G., {Leisenring}, J., {Wagner}, K.~R., {Mullin}, C., {Dong}, R.,
  {Greene}, T., {Johnstone}, D., {Meyer}, M.~R., {Wolff}, S.~G., {Beichman},
  C., {Boyer}, M., {Horner}, S., {Hodapp}, K., {Kelly}, D., {McCarthy}, D.,
  {Roellig}, T., {Rieke}, G., {Rieke}, M., {Stansberry}, J., and {Young}, E.,
  ``{JWST/NIRCam Imaging of Young Stellar Objects. II. Deep Constraints on
  Giant Planets and a Planet Candidate Outside of the Spiral Disk Around SAO
  206462},'' {\em \aj}~{\bf 167},  182 (Apr. 2024).

\bibitem{2023NatAs...7.1208W}
{Wagner}, K., {Stone}, J., {Skemer}, A., {Ertel}, S., {Dong}, R., {Apai}, D.,
  {Spalding}, E., {Leisenring}, J., {Sitko}, M., {Kratter}, K., {Barman}, T.,
  {Marley}, M., {Miles}, B., {Boccaletti}, A., {Assani}, K., {Bayyari}, A.,
  {Uyama}, T., {Woodward}, C.~E., {Hinz}, P., {Briesemeister}, Z., {Lawson},
  K., {M{\'e}nard}, F., {Pantin}, E., {Russell}, R.~W., {Skrutskie}, M., and
  {Wisniewski}, J., ``{Direct images and spectroscopy of a giant protoplanet
  driving spiral arms in MWC 758.},'' {\em Nature Astronomy}~{\bf 7},
  1208--1217 (Oct. 2023).

\bibitem{2013Sci...339.1398K}
{Konopacky}, Q.~M., {Barman}, T.~S., {Macintosh}, B.~A., and {Marois}, C.,
  ``{Detection of Carbon Monoxide and Water Absorption Lines in an Exoplanet
  Atmosphere},'' {\em Science}~{\bf 339},  1398--1401 (Mar. 2013).

\bibitem{2017AJ....154..218N}
{Nielsen}, E.~L., {Rosa}, R. J.~D., {Rameau}, J., {Wang}, J.~J., {Esposito},
  T.~M., {Millar-Blanchaer}, M.~A., {Marois}, C., {Vigan}, A., {Ammons}, S.~M.,
  {Artigau}, E., {Bailey}, V.~P., {Blunt}, S., {Bulger}, J., {Chilcote}, J.,
  {Cotten}, T., {Doyon}, R., {Duch{\^e}ne}, G., {Fabrycky}, D., {Fitzgerald},
  M.~P., {Follette}, K.~B., {Gerard}, B.~L., {Goodsell}, S.~J., {Graham},
  J.~R., {Greenbaum}, A.~Z., {Hibon}, P., {Hinkley}, S., {Hung}, L.-W.,
  {Ingraham}, P., {Jensen-Clem}, R., {Kalas}, P., {Konopacky}, Q., {Larkin},
  J.~E., {Macintosh}, B., {Maire}, J., {Marchis}, F., {Metchev}, S.,
  {Morzinski}, K.~M., {Murray-Clay}, R.~A., {Oppenheimer}, R., {Palmer}, D.,
  {Patience}, J., {Perrin}, M., {Poyneer}, L., {Pueyo}, L., {Rafikov}, R.~R.,
  {Rajan}, A., {Rantakyr{\"o}}, F.~T., {Ruffio}, J.-B., {Savransky}, D.,
  {Schneider}, A.~C., {Sivaramakrishnan}, A., {Song}, I., {Soummer}, R.,
  {Thomas}, S., {Wallace}, J.~K., {Ward-Duong}, K., {Wiktorowicz}, S., and
  {Wolff}, S., ``{Evidence That the Directly Imaged Planet HD 131399 Ab Is a
  Background Star},'' {\em \aj}~{\bf 154},  218 (Dec. 2017).

\bibitem{Beichman_2010}
Beichman, C.~A., Krist, J., Trauger, J.~T., Greene, T., Oppenheimer, B.,
  Sivaramakrishnan, A., Doyon, R., Boccaletti, A., Barman, T.~S., and Rieke,
  M., ``Imaging young giant planets from ground and space,'' {\em Publications
  of the Astronomical Society of the Pacific}~{\bf 122},  162–200 (Feb.
  2010).

\bibitem{2018arXiv180303730B}
{Beichman}, C.~A. and {Greene}, T.~P., ``{A White Paper Submitted to The
  National Academy of Science's Committee on Exoplanet Science Strategy:
  Observing Exoplanets with the James Webb Space Telescope},'' {\em arXiv
  e-prints} ,  arXiv:1803.03730 (Mar. 2018).

\bibitem{2005SPIE.5905..185G}
{Green}, J.~J., {Beichman}, C., {Basinger}, S.~A., {Horner}, S., {Meyer}, M.,
  {Redding}, D.~C., {Rieke}, M., and {Trauger}, J.~T., ``{High contrast imaging
  with the JWST NIRCAM coronagraph},'' in [{\em Techniques and Instrumentation
  for Detection of Exoplanets II}{\nolinebreak\hspace{0.1em}]},  {Coulter},
  D.~R., ed., {\em Society of Photo-Optical Instrumentation Engineers (SPIE)
  Conference Series} {\bf 5905},  185--195 (Aug. 2005).

\bibitem{2021MNRAS.501.1999C}
{Carter}, A.~L., {Hinkley}, S., {Bonavita}, M., {Phillips}, M.~W., {Girard},
  J.~H., {Perrin}, M., {Pueyo}, L., {Vigan}, A., {Gagn{\'e}}, J., and {Skemer},
  A. J.~I., ``{Direct imaging of sub-Jupiter mass exoplanets with James Webb
  Space Telescope coronagraphy},'' {\em \mnras}~{\bf 501},  1999--2016 (Feb.
  2021).

\bibitem{2024CRPhy..24S.133G}
{Galicher}, R. and {Mazoyer}, J., ``{Imaging exoplanets with coronagraphic
  instruments},'' {\em Comptes Rendus Physique}~{\bf 24},  133 (Jan. 2024).

\bibitem{2010SPIE.7731E..3JK}
{Krist}, J.~E., {Balasubramanian}, K., {Muller}, R.~E., {Shaklan}, S.~B.,
  {Kelly}, D.~M., {Wilson}, D.~W., {Beichman}, C.~A., {Serabyn}, E., {Mao}, Y.,
  {Echternach}, P.~M., {Trauger}, J.~T., and {Liewer}, K.~M., ``{The
  JWST/NIRCam coronagraph flight occulters},'' in [{\em Space Telescopes and
  Instrumentation 2010: Optical, Infrared, and Millimeter
  Wave}{\nolinebreak\hspace{0.1em}]},  {Oschmann}, Jacobus~M., J., {Clampin},
  M.~C., and {MacEwen}, H.~A., eds., {\em Society of Photo-Optical
  Instrumentation Engineers (SPIE) Conference Series} {\bf 7731},  77313J (July
  2010).

\bibitem{2012AAS...21915502T}
{Trauger}, J.~T., {Moody}, D., {Gordon}, B., {Krist}, J., and {Mawet}, D., ``{A
  Hybrid Lyot Coronagraph for the Direct Imaging and Spectroscopy of Exoplanet
  Systems: Recent Laboratory Demonstrations and Prospects},'' in [{\em American
  Astronomical Society Meeting Abstracts \#219}{\nolinebreak\hspace{0.1em}]},
  {\em American Astronomical Society Meeting Abstracts} {\bf 219},  155.02
  (Jan. 2012).

\bibitem{2019AJ....158...36G}
{Gerard}, B.~L., {Marois}, C., {Currie}, T., {Brandt}, T., {Chilcote}, J.~K.,
  {Draper}, Z.~H., {Groff}, T., {Guyon}, O., {Hayashi}, M., {Jovanovic}, N.,
  {Knapp}, G.~R., {Kudo}, T., {Kwon}, J., {Lozi}, J., {Martinache}, F.,
  {McElwain}, M., {Tamura}, M., and {Uyama}, T., ``{A Chromaticity Analysis and
  PSF Subtraction Techniques for SCExAO/CHARIS Data},'' {\em \aj}~{\bf 158},
  36 (July 2019).

\bibitem{1985SPIE..556..270M}
{M{\"u}ller}, M. and {Weigelt}, G., ``{Roll deconvolution of Space Telescope
  data: inverse filtering of two speckle interferograms.},'' in [{\em
  International Conference on Speckle}{\nolinebreak\hspace{0.1em}]},
  {Arsenault}, H.~H., ed., {\em Society of Photo-Optical Instrumentation
  Engineers (SPIE) Conference Series} {\bf 556},  270--273 (Jan. 1985).

\bibitem{2016SPIE.9909E..7QC}
{Carolo}, E., {Vassallo}, D., {Farinato}, J., {Bergomi}, M., {Bonavita}, M.,
  {Carlotti}, A., {D'Orazi}, V., {Greggio}, D., {Magrin}, D., {Mesa}, D.,
  {Pinna}, E., {Puglisi}, A., {Stangalini}, M., {Verinaud}, C., and {Viotto},
  V., ``{A comparison between different coronagraphic data reduction
  techniques},'' in [{\em Adaptive Optics Systems
  V}{\nolinebreak\hspace{0.1em}]},  {Marchetti}, E., {Close}, L.~M., and
  {V{\'e}ran}, J.-P., eds., {\em Society of Photo-Optical Instrumentation
  Engineers (SPIE) Conference Series} {\bf 9909},  99097Q (July 2016).

\bibitem{2022SPIE12180E..0SH}
{Hinkley}, S., {Carter}, A.~L., {Ray}, S., {Biller}, B., {Skemer}, A.,
  {Choquet}, E., {Millar-Blanchaer}, M.~A., {Sallum}, S., {Miles}, B.,
  {Whiteford}, N., {Patapis}, P., {Perrin}, M., {Pueyo}, L., {Stapelfeldt}, K.,
  {Wang}, J., {Ward-Duong}, K., {Girard}, J.~H., {Hines}, D., {Kammerer}, J.,
  {Leisenring}, J., {Zhou}, Y., {Meyer}, M., {Liu}, M.~C., {Bonnefoy}, M.,
  {Petrus}, S., {Bonavita}, M., {Chauvin}, G., {Chen}, C., {Currie}, T.,
  {Hoch}, K. K.~H., {Lazzoni}, C., {Matthews}, E.~C., {McElwain}, M.,
  {Rebollido}, I., {Rickman}, E., {Schneider}, G., {Sivaramakrishnan}, A., and
  {Stone}, J.~M., ``{Direct imaging and spectroscopy of exoplanetary systems
  with the JWST early release science program},'' in [{\em Space Telescopes and
  Instrumentation 2022: Optical, Infrared, and Millimeter
  Wave}{\nolinebreak\hspace{0.1em}]},  {Coyle}, L.~E., {Matsuura}, S., and
  {Perrin}, M.~D., eds., {\em Society of Photo-Optical Instrumentation
  Engineers (SPIE) Conference Series} {\bf 12180},  121800S (Aug. 2022).

\bibitem{2024AJ....167..181W}
{Wagner}, K., {Leisenring}, J., {Cugno}, G., {Mullin}, C., {Dong}, R., {Wolff},
  S.~G., {Greene}, T., {Johnstone}, D., {Meyer}, M.~R., {Beichman}, C.,
  {Boyer}, M., {Horner}, S., {Hodapp}, K., {Kelly}, D., {McCarthy}, D.,
  {Roellig}, T., {Rieke}, G., {Rieke}, M., {Sitko}, M., {Stansberry}, J., and
  {Young}, E., ``{JWST/NIRCam Imaging of Young Stellar Objects. I. Constraints
  on Planets Exterior to the Spiral Disk Around MWC 758},'' {\em \aj}~{\bf
  167},  181 (Apr. 2024).

\bibitem{2024AJ....167..183M}
{Mullin}, C., {Dong}, R., {Leisenring}, J., {Cugno}, G., {Greene}, T.,
  {Johnstone}, D., {Meyer}, M.~R., {Wagner}, K.~R., {Wolff}, S.~G., {Boyer},
  M., {Horner}, S., {Hodapp}, K., {McCarthy}, D., {Rieke}, G., {Rieke}, M., and
  {Young}, E., ``{JWST/NIRCam Imaging of Young Stellar Objects. III. Detailed
  Imaging of the Nebular Environment around the HL Tau Disk},'' {\em \aj}~{\bf
  167},  183 (Apr. 2024).

\bibitem{2019AJ....157..118R}
{Ruane}, G., {Ngo}, H., {Mawet}, D., {Absil}, O., {Choquet}, {\'E}., {Cook},
  T., {Gomez Gonzalez}, C., {Huby}, E., {Matthews}, K., {Meshkat}, T.,
  {Reggiani}, M., {Serabyn}, E., {Wallack}, N., and {Xuan}, W.~J., ``{Reference
  Star Differential Imaging of Close-in Companions and Circumstellar Disks with
  the NIRC2 Vortex Coronagraph at the W. M. Keck Observatory},'' {\em \aj}~{\bf
  157},  118 (Mar. 2019).

\bibitem{2022SPIE12180E..3QG}
{Girard}, J.~H., {Leisenring}, J., {Kammerer}, J., {Gennaro}, M., {Rieke}, M.,
  {Stansberry}, J., {Rest}, A., {Egami}, E., {Sunnquist}, B., {Boyer}, M.,
  {Canipe}, A., {Correnti}, M., {Hilbert}, B., {Perrin}, M.~D., {Pueyo}, L.,
  {Soummer}, R., {Allen}, M., {Bushouse}, H., {Aguilar}, J., {Brooks}, B.,
  {Coe}, D., {DiFelice}, A., {Golimowski}, D., {Hartig}, G., {Hines}, D.~C.,
  {Koekemoer}, A., {Nickson}, B., {Nikolov}, N., {Kozhurina-Platais}, V.,
  {Pirzkal}, N., {Robberto}, M., {Sivaramakrishnan}, A., {Sohn}, S.~T.,
  {Telfer}, R., {Wu}, C.~R., {Beatty}, T., {Florian}, M., {Hainline}, K.,
  {Kelly}, D., {Misselt}, K., {Schlawin}, E., {Sun}, F., {Williams}, C.,
  {Willmer}, C., {Stark}, C., {Ygouf}, M., {Carter}, A., {Beichman}, C.,
  {Greene}, T.~P., {Roellig}, T., {Krist}, J., {Adams Redai}, J., {Wang}, J.,
  {Clark}, C.~R., {Lewis}, D., and {Ferry}, M., ``{JWST/NIRCam coronagraphy:
  commissioning and first on-sky results},'' in [{\em Space Telescopes and
  Instrumentation 2022: Optical, Infrared, and Millimeter
  Wave}{\nolinebreak\hspace{0.1em}]},  {Coyle}, L.~E., {Matsuura}, S., and
  {Perrin}, M.~D., eds., {\em Society of Photo-Optical Instrumentation
  Engineers (SPIE) Conference Series} {\bf 12180},  121803Q (Aug. 2022).

\bibitem{2023PASP..135d8001R}
{Rigby}, J., {Perrin}, M., {McElwain}, M., {Kimble}, R., {Friedman}, S.,
  {Lallo}, M., {Doyon}, R., {Feinberg}, L., {Ferruit}, P., {Glasse}, A.,
  {Rieke}, M., {Rieke}, G., {Wright}, G., {Willott}, C., {Colon}, K., {Milam},
  S., {Neff}, S., {Stark}, C., {Valenti}, J., {Abell}, J., {Abney}, F.,
  {Abul-Huda}, Y., {Acton}, D.~S., {Adams}, E., {Adler}, D., {Aguilar}, J.,
  {Ahmed}, N., {Albert}, L., {Alberts}, S., {Aldridge}, D., {Allen}, M.,
  {Altenburg}, M., {{\'A}lvarez-M{\'a}rquez}, J., {Alves de Oliveira}, C.,
  {Andersen}, G., {Anderson}, H., {Anderson}, S., {Argyriou}, I., {Armstrong},
  A., {Arribas}, S., {Artigau}, E., {Arvai}, A., {Atkinson}, C., {Bacon}, G.,
  {Bair}, T., {Banks}, K., {Barrientes}, J., {Barringer}, B., {Bartosik}, P.,
  {Bast}, W., {Baudoz}, P., {Beatty}, T., {Bechtold}, K., {Beck}, T.,
  {Bergeron}, E., {Bergkoetter}, M., {Bhatawdekar}, R., {Birkmann}, S.,
  {Blazek}, R., {Blome}, C., {Boccaletti}, A., {B{\"o}ker}, T., {Boia}, J.,
  {Bonaventura}, N., {Bond}, N., {Bosley}, K., {Boucarut}, R., {Bourque}, M.,
  {Bouwman}, J., {Bower}, G., {Bowers}, C., {Boyer}, M., {Bradley}, L.,
  {Brady}, G., {Braun}, H., {Breda}, D., {Bresnahan}, P., {Bright}, S.,
  {Britt}, C., {Bromenschenkel}, A., {Brooks}, B., {Brooks}, K., {Brown}, B.,
  {Brown}, M., {Brown}, P., {Bunker}, A., {Burger}, M., {Bushouse}, H., {Cale},
  S., {Cameron}, A., {Cameron}, P., {Canipe}, A., {Caplinger}, J., {Caputo},
  F., {Cara}, M., {Carey}, L., {Carniani}, S., {Carrasquilla}, M.,
  {Carruthers}, M., {Case}, M., {Catherine}, R., {Chance}, D., {Chapman}, G.,
  {Charlot}, S., {Charlow}, B., {Chayer}, P., {Chen}, B., {Cherinka}, B.,
  {Chichester}, S., {Chilton}, Z., {Chonis}, T., {Clampin}, M., {Clark}, C.,
  {Clark}, K., {Coe}, D., {Coleman}, B., {Comber}, B., {Comeau}, T.,
  {Connolly}, D., {Cooper}, J., {Cooper}, R., {Coppock}, E., {Correnti}, M.,
  {Cossou}, C., {Coulais}, A., {Coyle}, L., {Cracraft}, M., {Curti}, M.,
  {Cuturic}, S., {Davis}, K., {Davis}, M., {Dean}, B., {DeLisa}, A.,
  {deMeester}, W., {Dencheva}, N., {Dencheva}, N., {DePasquale}, J.,
  {Deschenes}, J., {Hunor Detre}, {\"O}., {Diaz}, R., {Dicken}, D., {DiFelice},
  A., {Dillman}, M., {Dixon}, W., {Doggett}, J., {Donaldson}, T., {Douglas},
  R., {DuPrie}, K., {Dupuis}, J., {Durning}, J., {Easmin}, N., {Eck}, W.,
  {Edeani}, C., {Egami}, E., {Ehrenwinkler}, R., {Eisenhamer}, J.,
  {Eisenhower}, M., {Elie}, M., {Elliott}, J., {Elliott}, K., {Ellis}, T.,
  {Engesser}, M., {Espinoza}, N., {Etienne}, O., {Etxaluze}, M., {Falini}, P.,
  {Feeney}, M., {Ferry}, M., {Filippazzo}, J., {Fincham}, B., {Fix}, M.,
  {Flagey}, N., {Florian}, M., {Flynn}, J., {Fontanella}, E., {Ford}, T.,
  {Forshay}, P., {Fox}, O., {Franz}, D., {Fu}, H., {Fullerton}, A., {Galkin},
  S., {Galyer}, A., {Garc{\'\i}a Mar{\'\i}n}, M., {Gardner}, J.~P., {Gardner},
  L., {Garland}, D., {Garrett}, B., {Gasman}, D., {Gaspar}, A., {Gaudreau}, D.,
  {Gauthier}, P., {Geers}, V., {Geithner}, P., {Gennaro}, M., {Giardino}, G.,
  {Girard}, J., {Giuliano}, M., {Glassmire}, K., {Glauser}, A., {Glazer}, S.,
  {Godfrey}, J., {Golimowski}, D., {Gollnitz}, D., {Gong}, F., {Gonzaga}, S.,
  {Gordon}, M., {Gordon}, K., {Goudfrooij}, P., {Greene}, T., {Greenhouse}, M.,
  {Grimaldi}, S., {Groebner}, A., {Grundy}, T., {Guillard}, P., {Gutman}, I.,
  {Ha}, K.~Q., {Haderlein}, P., {Hagedorn}, A., {Hainline}, K., {Haley}, C.,
  {Hami}, M., {Hamilton}, F., {Hammel}, H., {Hansen}, C., {Harkins}, T.,
  {Harr}, M., {Hart}, J., {Hart}, Q., {Hartig}, G., {Hashimoto}, R., {Haskins},
  S., {Hathaway}, W., {Havey}, K., {Hayden}, B., {Hecht}, K., {Heller-Boyer},
  C., {Henriques}, C., {Henry}, A., {Hermann}, K., {Hernandez}, S., {Hesman},
  B., {Hicks}, B., {Hilbert}, B., {Hines}, D., {Hoffman}, M., {Holfeltz}, S.,
  {Holler}, B.~J., {Hoppa}, J., {Hott}, K., {Howard}, J.~M., {Howard}, R.,
  {Hunter}, A., {Hunter}, D., {Hurst}, B., {Husemann}, B., {Hustak}, L.,
  {Ilinca Ignat}, L., {Illingworth}, G., {Irish}, S., {Jackson}, W., {Jahromi},
  A., {Jakobsen}, P., {James}, L., {James}, B., {Januszewski}, W., {Jenkins},
  A., {Jirdeh}, H., {Johnson}, P., {Johnson}, T., {Jones}, V., {Jones}, R.,
  {Jones}, D., {Jones}, O., {Jordan}, I., {Jordan}, M., {Jurczyk}, S.,
  {Jurling}, A., {Kaleida}, C., {Kalmanson}, P., {Kammerer}, J., {Kang}, H.,
  {Kao}, S.-H., {Karakla}, D., {Kavanagh}, P., {Kelly}, D., {Kendrew}, S.,
  {Kennedy}, H., {Kenny}, D., {Keski-kuha}, R., {Keyes}, C., {Kidwell}, R.,
  {Kinzel}, W., {Kirk}, J., {Kirkpatrick}, M., {Kirshenblat}, D., {Klaassen},
  P., {Knapp}, B., {Knight}, J.~S., {Knollenberg}, P., {Koehler}, R.,
  {Koekemoer}, A., {Kovacs}, A., {Kulp}, T., {Kumari}, N., {Kyprianou}, M., {La
  Massa}, S., {Labador}, A., {Labiano}, A., {Lagage}, P.-O., {Lajoie}, C.-P.,
  {Lallo}, M., {Lam}, M., {Lamb}, T., {Lambros}, S., {Lampenfield}, R.,
  {Langston}, J., {Larson}, K., {Law}, D., {Lawrence}, J., {Lee}, D.,
  {Leisenring}, J., {Lepo}, K., {Leveille}, M., {Levenson}, N., {Levine}, M.,
  {Levy}, Z., {Lewis}, D., {Lewis}, H., {Libralato}, M., {Lightsey}, P.,
  {Link}, M., {Liu}, L., {Lo}, A., {Lockwood}, A., {Logue}, R., {Long}, C.,
  {Long}, D., {Loomis}, C., {Lopez-Caniego}, M., {Lorenzo Alvarez}, J.,
  {Love-Pruitt}, J., {Lucy}, A., {Luetzgendorf}, N., {Maghami}, P., {Maiolino},
  R., {Major}, M., {Malla}, S., {Malumuth}, E., {Manjavacas}, E., {Mannfolk},
  C., {Marrione}, A., {Marston}, A., {Martel}, A., {Maschmann}, M., {Masci},
  G., {Masciarelli}, M., {Maszkiewicz}, M., {Mather}, J., {McKenzie}, K.,
  {McLean}, B., {McMaster}, M., {Melbourne}, K., {Mel{\'e}ndez}, M., {Menzel},
  M., {Merz}, K., {Meyett}, M., {Meza}, L., {Miskey}, C., {Misselt}, K.,
  {Moller}, C., {Morrison}, J., {Morse}, E., {Moseley}, H., {Mosier}, G.,
  {Mountain}, M., {Mueckay}, J., {Mueller}, M., {Mullally}, S., {Murphy}, J.,
  {Murray}, K., {Murray}, C., {Mustelier}, D., {Muzerolle}, J., {Mycroft}, M.,
  {Myers}, R., {Myrick}, K., {Nanavati}, S., {Nance}, E., {Nayak}, O.,
  {Naylor}, B., {Nelan}, E., {Nickson}, B., {Nielson}, A., {Nieto-Santisteban},
  M., {Nikolov}, N., {Noriega-Crespo}, A., {O'Shaughnessy}, B., {O'Sullivan},
  B., {Ochs}, W., {Ogle}, P., {Oleszczuk}, B., {Olmsted}, J., {Osborne}, S.,
  {Ottens}, R., {Owens}, B., {Pacifici}, C., {Pagan}, A., {Page}, J., {Park},
  S., {Parrish}, K., {Patapis}, P., {Paul}, L., {Pauly}, T., {Pavlovsky}, C.,
  {Pedder}, A., {Peek}, M., {Pena-Guerrero}, M., {Penanen}, K., {Perez}, Y.,
  {Perna}, M., {Perriello}, B., {Phillips}, K., {Pietraszkiewicz}, M.,
  {Pinaud}, J.-P., {Pirzkal}, N., {Pitman}, J., {Piwowar}, A., {Platais}, V.,
  {Player}, D., {Plesha}, R., {Pollizi}, J., {Polster}, E., {Pontoppidan}, K.,
  {Porterfield}, B., {Proffitt}, C., {Pueyo}, L., {Pulliam}, C., {Quirt}, B.,
  {Quispe Neira}, I., {Ramos Alarcon}, R., {Ramsay}, L., {Rapp}, G., {Rapp},
  R., {Rauscher}, B., {Ravindranath}, S., {Rawle}, T., {Regan}, M., {Reichard},
  T.~A., {Reis}, C., {Ressler}, M.~E., {Rest}, A., {Reynolds}, P., {Rhue}, T.,
  {Richon}, K., {Rickman}, E., {Ridgaway}, M., {Ritchie}, C., {Rix}, H.-W.,
  {Robberto}, M., {Robinson}, G., {Robinson}, M., {Robinson}, O., {Rock}, F.,
  {Rodriguez}, D., {Rodriguez Del Pino}, B., {Roellig}, T., {Rohrbach}, S.,
  {Roman}, A., {Romelfanger}, F., {Rose}, P., {Roteliuk}, A., {Roth}, M.,
  {Rothwell}, B., {Rowlands}, N., {Roy}, A., {Royer}, P., {Royle}, P., {Rui},
  C., {Rumler}, P., {Runnels}, J., {Russ}, M., {Rustamkulov}, Z., {Ryden}, G.,
  {Ryer}, H., {Sabata}, M., {Sabatke}, D., {Sabbi}, E., {Samuelson}, B.,
  {Sapp}, B., {Sappington}, B., {Sargent}, B., {Sauer}, A., {Scheithauer}, S.,
  {Schlawin}, E., {Schlitz}, J., {Schmitz}, T., {Schneider}, A., {Schreiber},
  J., {Schulze}, V., {Schwab}, R., {Scott}, J., {Sembach}, K., {Shanahan}, C.,
  {Shaughnessy}, B., {Shaw}, R., {Shawger}, N., {Shay}, C., {Sheehan}, E.,
  {Shen}, S., {Sherman}, A., {Shiao}, B., {Shih}, H.-Y., {Shivaei}, I.,
  {Sienkiewicz}, M., {Sing}, D., {Sirianni}, M., {Sivaramakrishnan}, A.,
  {Skipper}, J., {Sloan}, G.~C., {Slocum}, C., {Slowinski}, S., {Smith}, E.,
  {Smith}, E., {Smith}, D., {Smith}, C., {Snyder}, G., {Soh}, W., {Sohn},
  S.~T., {Soto}, C., {Spencer}, R., {Stallcup}, S., {Stansberry}, J., {Starr},
  C., {Starr}, E., {Stewart}, A., {Stiavelli}, M., {Straughn}, A.,
  {Strickland}, D., {Stys}, J., {Summers}, F., {Sun}, F., {Sunnquist}, B.,
  {Swade}, D., {Swam}, M., {Swaters}, R., {Swoish}, R., {Taylor}, J.~M.,
  {Taylor}, R., {Te Plate}, M., {Tea}, M., {Teague}, K., {Telfer}, R., {Temim},
  T., {Thatte}, D., {Thompson}, C., {Thompson}, L., {Thomson}, S., {Tikkanen},
  T., {Tippet}, W., {Todd}, C., {Toolan}, S., {Tran}, H., {Trejo}, E.,
  {Truong}, J., {Tsukamoto}, C., {Tustain}, S., {Tyra}, H., {Ubeda}, L.,
  {Underwood}, K., {Uzzo}, M., {Van Campen}, J., {Vandal}, T., {Vandenbussche},
  B., {Vila}, B., {Volk}, K., {Wahlgren}, G., {Waldman}, M., {Walker}, C.,
  {Wander}, M., {Warfield}, C., {Warner}, G., {Wasiak}, M., {Watkins}, M.,
  {Weaver}, A., {Weilert}, M., {Weiser}, N., {Weiss}, B., {Weissman}, S.,
  {Welty}, A., {West}, G., {Wheate}, L., {Wheatley}, E., {Wheeler}, T.,
  {White}, R., {Whiteaker}, K., {Whitehouse}, P., {Whiteleather}, J.,
  {Whitman}, W., {Williams}, C., {Willmer}, C., {Willoughby}, S., {Wilson}, A.,
  {Wirth}, G., {Wislowski}, E., {Wolf}, E., {Wolfe}, D., {Wolff}, S.,
  {Workman}, B., {Wright}, R., {Wu}, C., {Wu}, R., {Wymer}, K., {Yates}, K.,
  {Yeager}, C., {Yeates}, J., {Yerger}, E., {Yoon}, J., {Young}, A., {Yu}, S.,
  {Zak}, D., {Zeidler}, P., {Zhou}, J., {Zielinski}, T., {Zincke}, C., and
  {Zonak}, S., ``{The Science Performance of JWST as Characterized in
  Commissioning},'' {\em \pasp}~{\bf 135},  048001 (Apr. 2023).

\bibitem{2022A&A...666A..32X}
{Xie}, C., {Choquet}, E., {Vigan}, A., {Cantalloube}, F., {Benisty}, M.,
  {Boccaletti}, A., {Bonnefoy}, M., {Desgrange}, C., {Garufi}, A., {Girard},
  J., {Hagelberg}, J., {Janson}, M., {Kenworthy}, M., {Lagrange}, A.-M.,
  {Langlois}, M., {Menard}, F., and {Zurlo}, A., ``{Reference-star differential
  imaging on SPHERE/IRDIS},'' {\em \aap}~{\bf 666},  A32 (Oct. 2022).

\bibitem{2024jwst.prop.6122B}
{Bowens-Rubin}, R., {Limbach}, M.~A., {Carter}, A., {Ertel}, S., {Girard}, J.,
  {Hinz}, P.~M., {Matthews}, E.~C., {Morley}, C., {Mukherjee}, S., {Salama},
  M., {Vanderburg}, A., and {Wagner}, K., ``{Cool kids on the block: The direct
  detection of cold ice giants and gas giants orbiting young low-mass
  neighbors}.'' JWST Proposal. Cycle 3, ID. \#6122 (Feb. 2024).

\bibitem{2024jwst.prop.5835C}
{Carter}, A., {Absil}, O., {Balmer}, W., {Biller}, B., {Bogat}, E., {Bonavita},
  M., {Booth}, M., {Bowens-Rubin}, R., {Bowler}, B., {Calissendorff}, P.,
  {Chauvin}, G., {Fontanive}, C., {Franson}, K., {Gagne}, J., {Girard}, J.,
  {Hinkley}, S., {Hoch}, K. K.~W., {Kammerer}, J., {Kennedy}, G., {Leisenring},
  J.~M., {Li}, Y., {Limbach}, M.~A., {Liu}, M.~C., {Macintosh}, B.~A.,
  {Matthews}, E.~C., {Meyer}, M.~R., {Millar-Blanchaer}, M.~A., {Morley}, C.,
  {Perrin}, M., {Pueyo}, L., {Ray}, S., {Rebollido}, I., {Rickman}, E.,
  {Skemer}, A., {Wang}, J.~J., {Ward-Duong}, K., and {Whiteford}, N., ``{Into
  The Spotlight: Unveiling Wide-Separation Sub-Jupiters for Future JWST
  Characterization}.'' JWST Proposal. Cycle 3, ID. \#5835 (Feb. 2024).

\bibitem{2000Msngr..99...31M}
{M{\o}ller}, P., ``{Spectral PSF subtraction I: the SPSF look-up-table
  method},'' {\em The Messenger}~{\bf 99},  31--33 (Mar. 2000).

\bibitem{JWST2024}
Documentation, J.~U., ``Jwst high-contrast imaging proposal planning.''
  \url{https://jwst-docs.stsci.edu/methods-and-roadmaps/jwst-high-contrast-imaging/jwst-high-contrast-imaging-proposal-planning/hci-psf-reference-stars}
  (2024).
\newblock Accessed: 2024-05-16.

\bibitem{2024ApJ...963L...2S}
{Sallum}, S., {Ray}, S., {Kammerer}, J., {Sivaramakrishnan}, A., {Cooper}, R.,
  {Greebaum}, A.~Z., {Thatte}, D., {De Furio}, M., {Factor}, S.~M., {Meyer},
  M.~R., {Stone}, J.~M., {Carter}, A., {Biller}, B., {Hinkley}, S., {Skemer},
  A., {Su{\'a}rez}, G., {Leisenring}, J.~M., {Perrin}, M.~D., {Kraus}, A.~L.,
  {Absil}, O., {Balmer}, W.~O., {Betti}, S.~K., {Boccaletti}, A., {Bonavita},
  M., {Bonnefoy}, M., {Booth}, M., {Bowler}, B.~P., {Briesemeister}, Z.~W.,
  {Bryan}, M.~L., {Calissendorff}, P., {Cantalloube}, F., {Chauvin}, G.,
  {Chen}, C.~H., {Choquet}, E., {Christiaens}, V., {Cugno}, G., {Currie}, T.,
  {Danielski}, C., {Dupuy}, T.~J., {Faherty}, J.~K., {Fitzgerald}, M.~P.,
  {Fortney}, J.~J., {Franson}, K., {Girard}, J.~H., {Grady}, C.~A., {Gonzales},
  E.~C., {Henning}, T., {Hines}, D.~C., {Hoch}, K. K.~W., {Hood}, C.~E.,
  {Howe}, A.~R., {Janson}, M., {Kalas}, P., {Kennedy}, G.~M., {Kenworthy},
  M.~A., {Kervella}, P., {Kitzmann}, D., {Kuzuhara}, M., {Lagrange}, A.-M.,
  {Lagage}, P.-O., {Lawson}, K., {Lazzoni}, C., {Lew}, B. W.~P., {Liu}, M.~C.,
  {Liu}, P., {Llop-Sayson}, J., {Lloyd}, J.~P., {Lueber}, A., {Macintosh}, B.,
  {Manjavacas}, E., {Marino}, S., {Marley}, M.~S., {Marois}, C., {Martinez},
  R.~A., {Matthews}, B.~C., {Matthews}, E.~C., {Mawet}, D., {Mazoyer}, J.,
  {McElwain}, M.~W., {Metchev}, S., {Miles}, B.~E., {Millar-Blanchaer}, M.~A.,
  {Molliere}, P., {Moran}, S.~E., {Morley}, C.~V., {Mukherjee}, S.,
  {Palma-Bifani}, P., {Pantin}, E., {Patapis}, P., {Petrus}, S., {Pueyo}, L.,
  {Quanz}, S.~P., {Quirrenbach}, A., {Rebollido}, I., {Redai}, J.~A., {Ren},
  B.~B., {Rickman}, E., {Samland}, M., {Sargent}, B.~A., {Schlieder}, J.~E.,
  {Schneider}, G., {Stapelfeldt}, K.~R., {Sutlieff}, B.~J., {Tamura}, M.,
  {Tan}, X., {Theissen}, C.~A., {Uyama}, T., {Vigan}, A., {Vasist}, M., {Vos},
  J.~M., {Wagner}, K., {Wang}, J.~J., {Ward-Duong}, K., {Whiteford}, N.,
  {Wolff}, S.~G., {Worthen}, K., {Wyatt}, M.~C., {Ygouf}, M., {Zhang}, X.,
  {Zhang}, K., {Zhang}, Z., {Zhou}, Y., and {Zurlo}, A., ``{The JWST Early
  Release Science Program for Direct Observations of Exoplanetary Systems. IV.
  NIRISS Aperture Masking Interferometry Performance and Lessons Learned},''
  {\em \apjl}~{\bf 963},  L2 (Mar. 2024).

\bibitem{2021SPIE11823E..0HC}
{Carter}, A.~L., {Skemer}, A. J.~I., {Danielski}, C., {Leisenring}, J., {Wang},
  J.~J., {Van Gorkom}, K., {York}, B., {Adams}, J., {Biller}, B., {Girard},
  J.~H., {Hinkley}, S., {Nickson}, B., {Perrin}, M., and {Pueyo}, L.,
  ``{Simulating JWST high contrast observations with PanCAKE},'' in [{\em
  Techniques and Instrumentation for Detection of Exoplanets
  X}{\nolinebreak\hspace{0.1em}]},  {Shaklan}, S.~B. and {Ruane}, G.~J., eds.,
  {\em Society of Photo-Optical Instrumentation Engineers (SPIE) Conference
  Series} {\bf 11823},  118230H (Sept. 2021).

\bibitem{here}
 (here).
\newblock Official Documentation Site for PanCAKE.

\bibitem{2012SPIE.8442E..3DP}
{Perrin}, M.~D., {Soummer}, R., {Elliott}, E.~M., {Lallo}, M.~D., and
  {Sivaramakrishnan}, A., ``{Simulating point spread functions for the James
  Webb Space Telescope with WebbPSF},'' in [{\em Space Telescopes and
  Instrumentation 2012: Optical, Infrared, and Millimeter
  Wave}{\nolinebreak\hspace{0.1em}]},  {Clampin}, M.~C., {Fazio}, G.~G.,
  {MacEwen}, H.~A., and {Oschmann}, Jacobus~M., J., eds., {\em Society of
  Photo-Optical Instrumentation Engineers (SPIE) Conference Series} {\bf 8442},
   84423D (Sept. 2012).

\bibitem{2015ascl.soft06001W}
{Wang}, J.~J., {Ruffio}, J.-B., {De Rosa}, R.~J., {Aguilar}, J., {Wolff},
  S.~G., and {Pueyo}, L., ``{pyKLIP: PSF Subtraction for Exoplanets and
  Disks}.'' Astrophysics Source Code Library, record ascl:1506.001 (June 2015).

\bibitem{klaus-stephenson_2024}
github page,
  ``caffeine-deprived/the-effect-of-off-axis-sources-on-jwst-nircam-coronagraphic-performance,''
  (2024).
\newblock GitHub page for custom scripts and PanCAKE simulations.

\bibitem{anaconda}
``Anaconda software distribution.'' \url{https://anaconda.com} (Nov. 2016).
\newblock Computer software. Vers. 2-2.4.0.

\bibitem{2023ApJ...951L..20C}
{Carter}, A.~L., {Hinkley}, S., {Kammerer}, J., {Skemer}, A., {Biller}, B.~A.,
  {Leisenring}, J.~M., {Millar-Blanchaer}, M.~A., {Petrus}, S., {Stone}, J.~M.,
  {Ward-Duong}, K., {Wang}, J.~J., {Girard}, J.~H., {Hines}, D.~C., {Perrin},
  M.~D., {Pueyo}, L., {Balmer}, W.~O., {Bonavita}, M., {Bonnefoy}, M.,
  {Chauvin}, G., {Choquet}, E., {Christiaens}, V., {Danielski}, C., {Kennedy},
  G.~M., {Matthews}, E.~C., {Miles}, B.~E., {Patapis}, P., {Ray}, S.,
  {Rickman}, E., {Sallum}, S., {Stapelfeldt}, K.~R., {Whiteford}, N., {Zhou},
  Y., {Absil}, O., {Boccaletti}, A., {Booth}, M., {Bowler}, B.~P., {Chen},
  C.~H., {Currie}, T., {Fortney}, J.~J., {Grady}, C.~A., {Greebaum}, A.~Z.,
  {Henning}, T., {Hoch}, K. K.~W., {Janson}, M., {Kalas}, P., {Kenworthy},
  M.~A., {Kervella}, P., {Kraus}, A.~L., {Lagage}, P.-O., {Liu}, M.~C.,
  {Macintosh}, B., {Marino}, S., {Marley}, M.~S., {Marois}, C., {Matthews},
  B.~C., {Mawet}, D., {McElwain}, M.~W., {Metchev}, S., {Meyer}, M.~R.,
  {Molliere}, P., {Moran}, S.~E., {Morley}, C.~V., {Mukherjee}, S., {Pantin},
  E., {Quirrenbach}, A., {Rebollido}, I., {Ren}, B.~B., {Schneider}, G.,
  {Vasist}, M., {Worthen}, K., {Wyatt}, M.~C., {Briesemeister}, Z.~W., {Bryan},
  M.~L., {Calissendorff}, P., {Cantalloube}, F., {Cugno}, G., {De Furio}, M.,
  {Dupuy}, T.~J., {Factor}, S.~M., {Faherty}, J.~K., {Fitzgerald}, M.~P.,
  {Franson}, K., {Gonzales}, E.~C., {Hood}, C.~E., {Howe}, A.~R., {Kuzuhara},
  M., {Lagrange}, A.-M., {Lawson}, K., {Lazzoni}, C., {Lew}, B. W.~P., {Liu},
  P., {Llop-Sayson}, J., {Lloyd}, J.~P., {Martinez}, R.~A., {Mazoyer}, J.,
  {Palma-Bifani}, P., {Quanz}, S.~P., {Redai}, J.~A., {Samland}, M.,
  {Schlieder}, J.~E., {Tamura}, M., {Tan}, X., {Uyama}, T., {Vigan}, A., {Vos},
  J.~M., {Wagner}, K., {Wolff}, S.~G., {Ygouf}, M., {Zhang}, X., {Zhang}, K.,
  and {Zhang}, Z., ``{The JWST Early Release Science Program for Direct
  Observations of Exoplanetary Systems I: High-contrast Imaging of the
  Exoplanet HIP 65426 b from 2 to 16 {\ensuremath{\mu}}m},'' {\em \apjl}~{\bf
  951},  L20 (July 2023).

\bibitem{2024jwst.prop.6012M}
{Millar-Blanchaer}, M.~A., {Altinier}, L., {Carter}, A., {Choquet}, E., {De
  Rosa}, R.~J., {Girard}, J., {Godoy}, N., {Hinkley}, S., {Leisenring}, J.,
  {Macintosh}, B.~A., {Pearce}, T., {Ray}, S., {Vancil}, C.~J., and {Vigan},
  A., ``{Finding the great sculptors: A Renaissance in Planet Disk Dynamics}.''
  JWST Proposal. Cycle 3, ID. \#6012 (Feb. 2024).

\bibitem{2023jwst.prop.4050C}
{Carter}, A., {Balmer}, W., {Biller}, B., {Bogat}, E., {Bonavita}, M.,
  {Bowler}, B., {Calissendorff}, P., {Fontanive}, C., {Franson}, K., {Gagne},
  J., {Girard}, J., {Hinkley}, S., {Hoch}, K. K.~W., {Kammerer}, J., {Kennedy},
  G., {Leisenring}, J.~M., {Macintosh}, B.~A., {Matthews}, E.~C., {Meyer},
  M.~R., {Millar-Blanchaer}, M.~A., {Morley}, C., {Perrin}, M., {Pueyo}, L.,
  {Ray}, S., {Rebollido}, I., {Rickman}, E., {Skemer}, A., and {Wang}, J.~J.,
  ``{Uncharted Worlds: Towards a Legacy of Direct Imaging of Sub-Jupiter Mass
  Exoplanets}.'' JWST Proposal. Cycle 2, ID. \#4050 (May 2023).

\bibitem{2023jwst.prop.3989H}
{Hinkley}, S., {Lazzoni}, C., {Marino}, S., {Carter}, A., {Kennedy}, G.,
  {Milli}, J., and {Ray}, S., ``{Spotting the Perturbers: A Coronagraphic
  Survey of Debris Disk Stars with Proper Motion Anomalies}.'' JWST Proposal.
  Cycle 2, ID. \#3989 (May 2023).

\bibitem{nelan_2005}
Nelan, E.,  [{\em JWST Science Instrument Target Acquisition
  Concepts}{\nolinebreak\hspace{0.1em}]} (2005).

\bibitem{fox_2021}
Fox, K., ``Nasa says webb’s excess fuel likely to extend its lifetime
  expectations – james webb space telescope,'' (2021).

\end{thebibliography}
\bibliographystyle{spiebib}

\appendix
\section{Additional Plots} \label{bonus}
The following two sets of `heatmap' plots are additional expansions to those found in section \ref{discussion} covering binary companions as a function of position angle at relative brightnesses of $10^{-4}$ and $10^{-5}$.
\begin{figure}[ht!]
    \includegraphics[width=16cm]{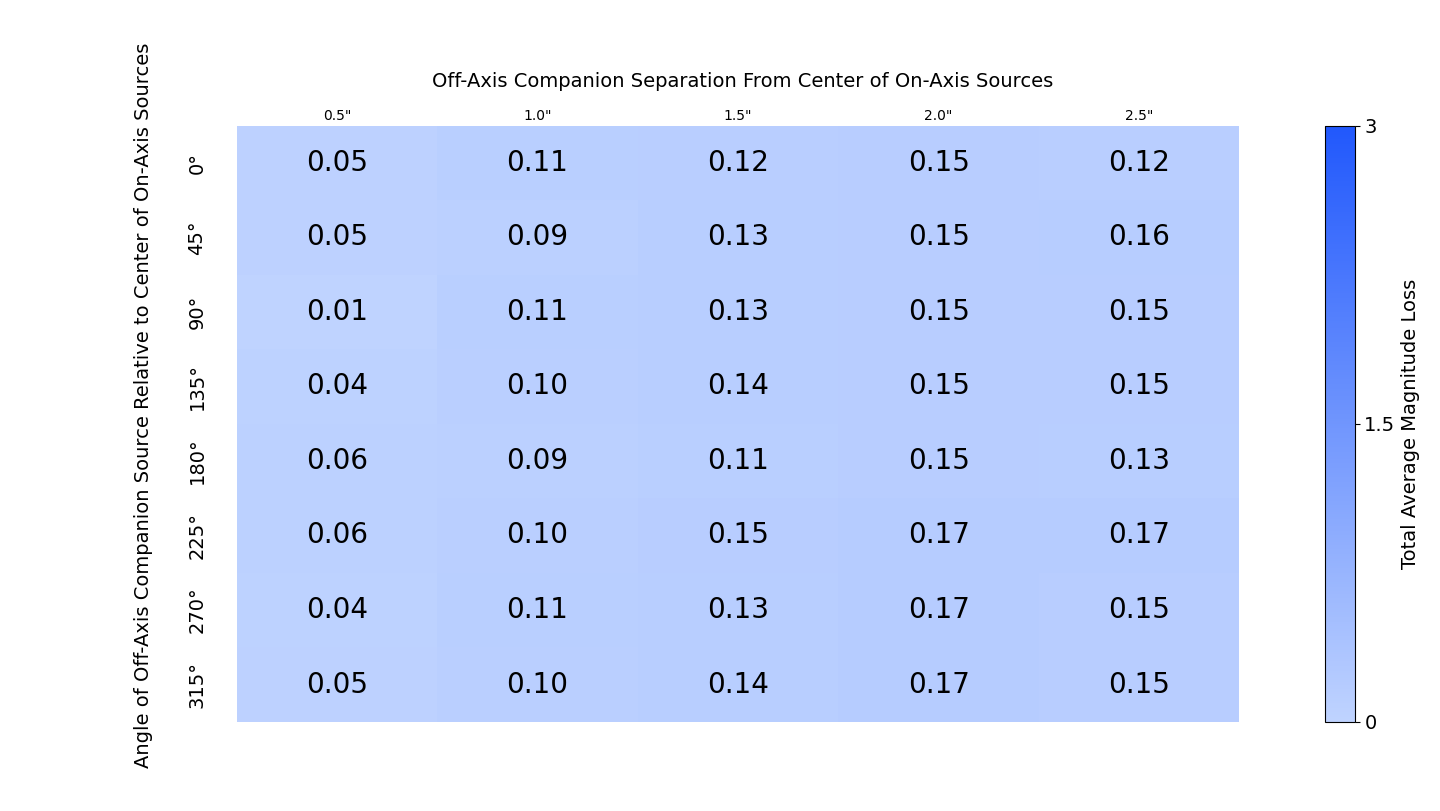}
    \includegraphics[width=16cm]{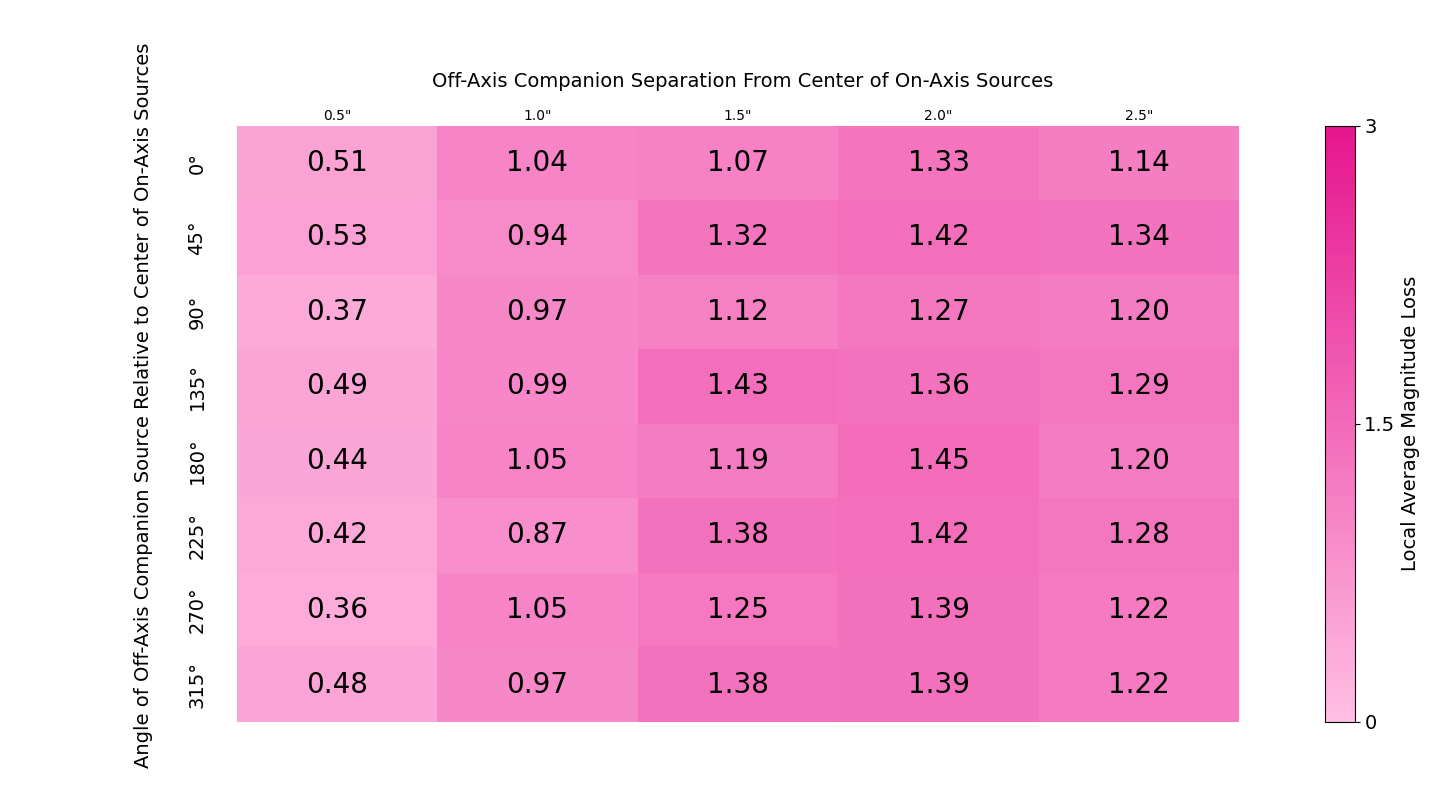}
    \centering
    \caption{Magnitude loss for varying binary companion position angles as a function of separation, with the binary companion relative brightness being $10^{-4}$ that of the on-axis source. The blue colored plot is total average magnitude sensitivity loss, the average loss over an entire image. The pink colored plot is local average magnitude sensitivity loss, or the average loss within 1" of the binary companion. All values for the local plot have an uncertainty of $\pm 0.059$; all values for the total average loss plot have an uncertainty of $\pm 0.032$.} \label{heatmap-04}
\end{figure}
\begin{figure}[ht!] 
    \includegraphics[width=16cm]{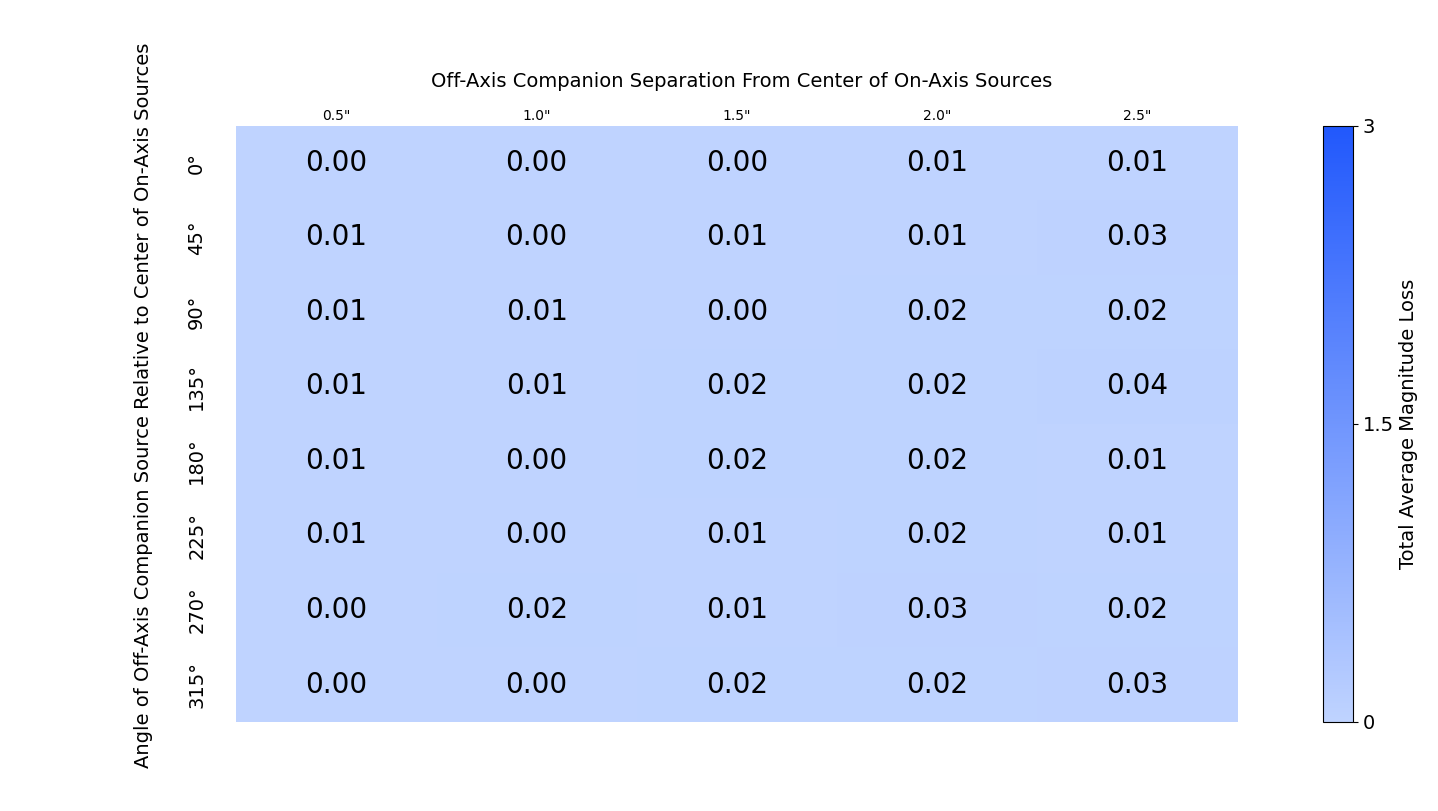}
    \includegraphics[width=16cm]{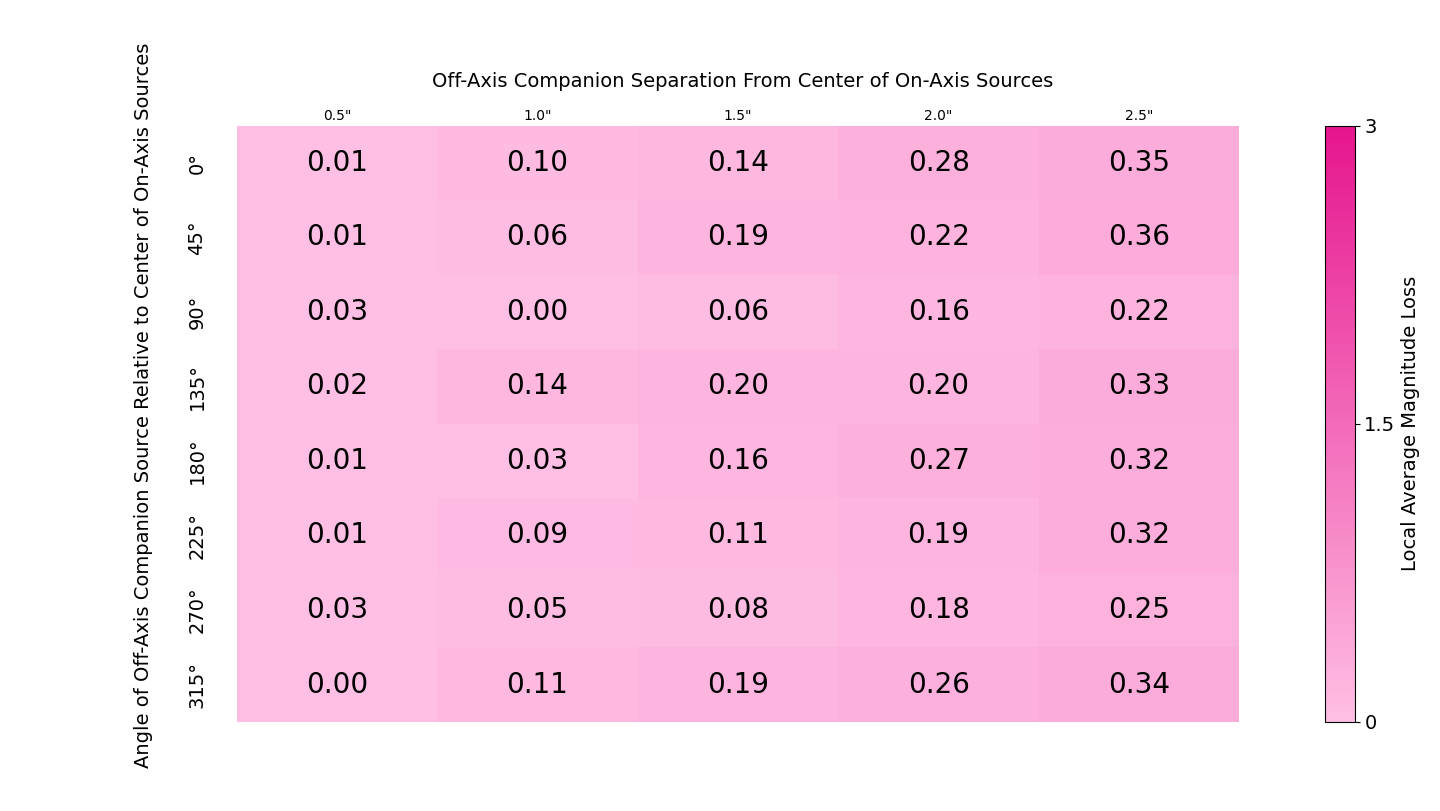}
    \centering
    \caption{Magnitude loss for varying binary companion position angles as a function of separation, with the binary companion relative brightness being $10^{-5}$ that of the on-axis source. The blue colored plot is total average magnitude sensitivity loss, the average loss over an entire image. The pink colored plot is local average magnitude sensitivity loss, or the average loss within 1" of the binary companion. All values for the local plot have an uncertainty of $\pm 0.059$; all values for the total average loss plot have an uncertainty of $\pm 0.032$.} \label{heatmap-05}
\end{figure}

\end{document}